\documentclass{article}
\usepackage{graphicx}
\usepackage{amssymb}
\usepackage{amsmath}
\usepackage{graphicx}
\usepackage{color}
\usepackage{listings}
\usepackage{tikz}
\usepackage{xcolor}
\usetikzlibrary{intersections, calc}

\newtheorem{prop}{Proposition}
\newtheorem{hodai}{Lemma}

\definecolor{Brown}{cmyk}{0,0.81,1,0.60}
\definecolor{OliveGreen}{cmyk}{0.64,0,0.95,0.40}
\definecolor{CadetBlue}{cmyk}{0.62,0.57,0.23,0}

\usepackage{graphicx}
\usepackage{amsmath}
\usepackage{amssymb}

\usepackage[round]{natbib}   
\title{Post-Selection Inference for Sparse Estimation\thanks{This manuscript is the English translation of an article originally published in Japanese in the Journal of the Japan Statistical Society, Volume 53, Issue 1, September 2023 (pages 139-167)}}
\author{Joe Suzuki\thanks{Graduate School of Engineering Science, Osaka University, 1-3 Machikaneyama-cho, Toyonaka, Osaka 560-8531, Japan \newline E-mail: prof.joe.suzuki@gmail.com}}

\begin{document}
\label{firstpage}

\definecolor{darkslateblue}{rgb}{0.28,0.24,0.55}
\definecolor{darkgreen}{rgb}{0.00,0.39,0.00}

\lstset{ 
  language=R,
  basicstyle={\small\ttfamily},
  stringstyle ={\ttfamily},
  commentstyle={\itshape\color{Brown}},
  identifierstyle={\ttfamily\color{darkslateblue}\bfseries}, 
  keywordstyle={\ttfamily\color{darkgreen}},
  breaklines=true,
  columns=[l]{fullflexible},
  lineskip=0mm,
  showstringspaces=false,
  keepspaces=true,
  frame=single, 
  numbers=left, 
  stepnumber=1, 
  numberstyle={\tiny}, 
} 

\maketitle

\subsubsection*{Abstract}
When the model is not known and parameter testing or interval estimation is conducted after model selection, it is necessary to consider selective inference. 
This paper discusses this issue in the context of sparse estimation. Firstly, we describe selective inference related to Lasso as per \cite{lee}, 
and then present polyhedra and truncated distributions when applying it to methods such as Forward Stepwise and LARS. Lastly, we discuss the Significance Test for Lasso by \cite{significant} and the Spacing Test for LARS by \cite{ryan_exact}. This paper serves as a review article.

\subsubsection*{Keywords}
post-selective inference, polyhedron, LARS, lasso, forward stepwise, significance test, spacing test.

\section{Introduction}

In this study, we explore the discrepancies that arise between conducting parameter tests and estimating confidence intervals when the model is known, and performing inference regarding the parameters after selecting a model. Furthermore, variables selected by performing model selection tend to be significant in the first place. The question then becomes: how can we eliminate this bias?

\begin{figure}
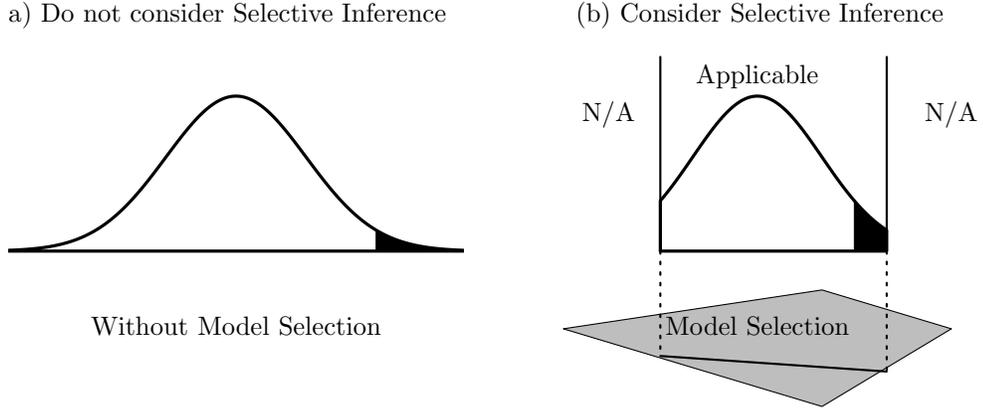

\centering
\begin{tabular}{cc}
\input{fig002_eng}&\input{fig001_eng}
\end{tabular}
\caption{Distribution of the test statistic under the null hypothesis. If the model is known, as in (a), it is permissible to perform regular statistical inference, considering $n$ samples $y$ move in $\mathbb{R}^n$. However, if a model $\hat{M}(y^{obs})$ is obtained from observed value $y^{obs}$ after conducting model selection, as in (b), statistical inference must be performed according to the distribution (truncated distribution) within the set of $y$ obtaining the same model $\{y\in \mathbb{R}^n|\hat{M}(y)=\hat{M}(y^{obs})\}$ (truncated region).\label{fig003}}
\end{figure}

This issue is known as Post-Selective Inference and has been specifically discussed, triggered by the works of \cite{berk} and \cite{lee}.
As seen in Figure \ref{fig003}, if the model is known, it is permissible to conduct regular parameter estimation, with $n$ samples $y$ moving in $\mathbb{R}^n$. However, if a model $\hat{M}(y^{obs})$ is obtained from the observed value $y^{obs}$ after conducting model selection, statistical inference must be performed according to the distribution (truncated distribution) within the set of $y$ obtaining the same model $\{y\in \mathbb{R}^n|\hat{M}(y)=\hat{M}(y^{obs})\}$ (truncated region). It could be said that the post-selective inference discussed in this paper pertains to this issue of the truncated region and distribution. Indeed, aside from that, it aligns with conventional statistical theory.

Although the special feature this time is "Sparse Estimation," the issue of selective inference is always a topic of discussion whenever there is a variable selection procedure, not limited to sparse estimation, but also in cases where information criteria are applied.

In sparse estimation, Lasso (least absolute shrinkage and selection operator, \citet{lasso}) is most commonly used. In the case of linear regression (assuming $n$ samples and $p$ variables), given $X \in \mathbb{R}^{n \times p}$, $y \in \mathbb{R}^n$, and a constant $\lambda > 0$, it is formulated as a problem to find $\beta = [\beta_1, \ldots, \beta_p]^\top \in \mathbb{R}^p$ that minimizes
\begin{equation}\label{eq1111}
\frac{1}{2}\|y-X\beta\|_2^2+\lambda \sum_{j=1}^p|\beta_j|.
\end{equation}
The larger the $\lambda$, the fewer non-zero components of $\beta$ (the fewer selected variables). Particularly, even when $n < p$, in other words, even when $X^\top X$ does not have an inverse matrix, a solution can be found\footnote{When $n<p$, it is not strongly convex, and the solution is not unique. However, uniqueness is guaranteed when samples arise according to a continuous distribution \citep{ryan_unique}.}. Furthermore, because (\ref{eq1111}) is a convex function of $\beta \in \mathbb{R}^p$, a solution that minimizes (\ref{eq1111}) can be efficiently found. Here, $\lambda$ is usually determined from $X, y$ using methods such as cross-validation.

Selective inference, along with sparse estimation, could be described as one of the specialties of Stanford Statistics, and many studies, starting with \cite{lee}, have been conducted in the context of sparse estimation. Since (\ref{eq1111}) includes an absolute value term, deriving confidence intervals was originally difficult. On the other hand, another sparse estimation method, SCAD (\cite{scad}), has a method of deriving confidence intervals based on oracle properties, but optimization in large-scale problems is difficult because the objective function is not convex. Therefore, there is a two-step estimation that performs variable selection with Lasso and derives confidence intervals using the selected variables, but inevitably, the variables selected by performing model selection tend to be significant.

Several explanatory articles regarding selective inference have already been published. This paper focuses on its theoretical handling and sequential model selection.

Understanding the essence of the work by \cite{lee} is crucial for selective inference. Starting with selective inference assuming linear regression (Section 2), while confirming the theoretical treatment of Lasso (Section 3), we derive the theory of polyhedra constructing the truncated distribution (Section 4). Although it includes some formal discussions, this part corresponds to the preparation of this paper.

Discussion of sequential model selection inevitably requires understanding the framework of Forward Stepwise (FS). Even in linear regression, if there are $p$ candidate explanatory variables, it is necessary to choose the optimal combination from $2^p$ combinations, making the theoretical analysis of variable selection difficult. Section 5 applies FS, which may not be optimal but selects variables top-down one by one, to perform analysis related to variable selection and residuals. LARS by Efron (\cite{efron}, Section 6) was also conceived by applying FS-like concepts to Lasso. It is a method that shows almost the same performance as Lasso and is considered easy to theoretically analyze.

The sequential model selection dealt with in this paper (Sections 7 and 8) will be a discussion based on LARS. It repeatedly tests whether the coefficient in front of each explanatory variable of linear regression should be 0 or not. At that time, testing whether the coefficient in front of the variable at each point is 0 or not is performed based on different distributions that depend on the variables already selected. This framework seems like a generalization of the concept of selective inference.

The author had a problem awareness that discussions on sequential selective inference, while important, are not well known. It may be a challenging problem that you cannot understand the essence unless you understand all of Stanford Statistics' products, such as selective inference, Lasso, FS, and LARS. In this paper, we explain them in order and reach the peak of Significance Test (\cite{significant}, Section 7) and Spacing Test (\cite{ryan_exact}, Section 8).

Section 9 summarizes the overall and mentions future prospects. To maintain self-containedness, a proof (provided an easy proof different from the original paper) was posted in the appendix.

Additionally, this paper provides source programs in R language (Table \ref{tab7}). Please utilize appropriately.

\begin{table}[h]
\centering
\caption{\label{tab7}Related source codes}
\begin{tabular}{l|l}
\hline
Compute truncated distribution & \verb|https://bayesnet.org/books_jp/?p=576| \\
Lasso (Figure \ref{fig1-7}) & \verb|https://bayesnet.org/books_jp/3-1.html| \\
LARS (Figure \ref{fig1-7}) & \verb|https://bayesnet.org/books_jp/3-4.html| \\
General sparse estimation & \verb|https://bayesnet.org/books_jp/?page_id=33| \\
\hline
\end{tabular}
\end{table}

\section{Selective Inference in Linear Regression}

Let us consider $n$ samples $(x_1,y_1),\ldots,(x_n,y_n) \in \mathbb{R}^p \times \mathbb{R}$, where $x_i = [x_{i,1},\ldots,x_{i,p}] \in \mathbb{R}^p$ (row vector\footnote{Vectors are typically represented as column vectors.}), $i=1,\ldots,n$. We aim to find $\beta_0, \beta_1, \ldots, \beta_p$ that minimize
$$
L := \sum_{i=1}^n \left( y_i - \beta_0 - \sum_{j=1}^p x_{i,j} \beta_j \right)^2
$$
through the method of least squares. Hereafter, we denote by $X$ the $n \times p$ matrix whose rows are $x_i$, and by $y$ the $n$-dimensional vector with components $y_i$. Without loss of generality, we assume that each column of $X$ and $y$ has been centered, and we exclude the intercept, thereby setting $\beta_0 = 0$, for further discussion.
The solution obtained by setting the derivative of $L$ with respect to $\beta = [\beta_1,\ldots,\beta_p]^\top$ to zero is denoted as $\hat{\beta}_1,\ldots,\hat{\beta}_p$.
Assume there exists some $\beta \in \mathbb{R}^p$ (the true parameter), such that
\[ y_i - \sum_{j=1}^p x_{i,j} \beta_j \]
for $i=1,\ldots,n$ are independent and normally distributed with mean 0 and known variance $\sigma^2$. Assuming $X^\top X$ is non-singular, the coefficients obtained through the least squares method, $\hat{\beta} = [\hat{\beta}_1,\ldots,\hat{\beta}_p]^\top$, are random variables and, omitting detailed derivation, can be expressed as
\begin{equation}\label{eq1}
\hat{\beta} = (X^\top X)^{-1} X^\top y \sim N(\beta, \sigma^2 (X^\top X)^{-1})
\end{equation}
where $\sim N(\mu,\Sigma)$ denotes a normal distribution in $p$ dimensions with mean vector $\mu$ and $p \times p$ covariance matrix $\Sigma$.

In linear regression, it is often assumed that among the $p$ explanatory variables, some are redundant and their parameter $\beta_j$ should be set to 0. We can identify redundant parameters by testing the null hypothesis $H_0: \beta_j = 0$ and alternative hypothesis $H_1: \beta_j \neq 0$ for each $j = 1, \ldots, p$. If we define
\[ RSS := \|y - X\hat{\beta}\|_2^2 \]
then
\[ \frac{RSS}{\sigma^2} \sim \chi^2_{n-p} \]
and using equation (\ref{eq1}), if we define $\hat{\sigma} := \sqrt{RSS/(n-p)}$, $B_j$ as the $j$-th diagonal element of $(X^\top X)^{-1}$, and $SE(\hat{\beta}_j) := \sqrt{B_j}\hat{\sigma}$, then
\[ t := \frac{\hat{\beta}_j - \beta_j}{SE(\hat{\beta}_j)} = \frac{(\hat{\beta}_j - \beta_j)/\sqrt{B_j}}{\sqrt{RSS/(n-p)}} \]
follows a $t$-distribution with $n-p$ degrees of freedom. Under the null hypothesis $\beta_j = 0$, if the observed value ${\hat{\beta}_j}/{SE(\hat{\beta}_j)}$ falls outside the $\alpha/2$ percentile of its null distribution for a significance level $0<\alpha<1$, we consider the variable significant and decide $\beta_j \neq 0$.

For variable selection, methods such as the stepwise method or Lasso are available. We assume the existence of a mapping
\[ \hat{M}: \mathbb{R}^n \rightarrow ({\rm the\ set\ of\ subsets\ of\ } \{1,\ldots,p\}) \]
where $\mathcal{P}(S)$ denotes the power set of $S$. When using these methods to identify the model $\hat{M}(y) = \{j \in \{1,\ldots,p\}|\hat{\beta}_j \neq 0\}$, parameters $\beta_j$ for which $\hat{\beta}_j \neq 0$, $j \in \hat{M}(y)$, often yield small $p$-values. Moreover, parameters that are not selected are presumed to be 0. Consequently, it may arise that parameter confidence intervals and testing should be conducted conditionally given the selected model $M = \hat{M}(y)$. Selective inference advocates for parameter confidence intervals and testing based on the conditional distribution given the selection and applies more broadly than just linear regression, being relevant for statistical estimation involving variable selection in general.

Returning to linear regression, let's define the confidence level as \(1-\alpha\) (\(\alpha > 0\)), and instead of defining a confidence interval \(C\) as
\[ {\mathbb P}\left({\beta}_j^{\hat{M}(y)}\in C\right)\geq 1-\alpha \]
it becomes defined under \(M=\hat{M}(y)\) as
\[ {\mathbb P}\left({\beta}^{M}_j\in C|M=\hat{M}(y)\right)\geq 1-\alpha \]
where we denote the true value of the $j$-th parameter assuming the model $M$ by ${\beta}_j^M$. Indeed, if the model $M$ differs, the parameter \(\beta^M\in {\mathbb R}^m\) (\(m:=|M|\)) also differs.

Here, writing \(X_M\in {\mathbb R}^{n\times m}\) and expressing the submatrix of \(X\) corresponding to the index set \(M\subseteq \{1,\ldots,p\}\), the estimate of ${\beta}^M_j$ is written as \(\hat{\beta}^M_j=e_j^\top(X_M^\top X_M)^{-1}X_M^\top y\). Here, let \(e_j\in {\mathbb R}^m\) be a vector where only the $j$-th component is 1 and the rest are 0. Below, denoting \(\eta=X_M(X_M^\top X_M)^{-1}e_j\in {\mathbb R}^{n}\) and assuming \(y^{obs}\) is observed as \(y\)\footnote{All of \(y_1,\ldots,y_n\) are observed. The fixed observed values are written as \(y^{obs}\in {\mathbb R}^n\) and distinguished from the random variable \(y\in {\mathbb R}^n\).}, we determine the confidence interval and \(p\) value of \(\beta_j^M\) and \(\hat{\beta}_j^M\) based on the conditional probability
\[ \eta^\top y\mid \left\{\hat{M}(y)=\hat{M}(y^{obs})\right\}. \]

At this moment, the confidence interval for \(\beta_j^M\) at a confidence level of \(1-\alpha\) \((0 < \alpha < 1)\) is established by employing \(L, U\) that satisfy
\[
\mathbb{P}\left(\eta^\top y^{\text{obs}}\leq \eta^\top y|\beta_j^M=L, \hat{M}(y)=\hat{M}(y^{\text{obs}})\right)=\frac{\alpha}{2}
\]
\[
\mathbb{P}\left(\eta^\top y\leq \eta^\top y^{\text{obs}}|\beta_j^M=U, \hat{M}(y)=\hat{M}(y^{\text{obs}})\right)=\frac{\alpha}{2}
\]
hence, it can be written as,
\begin{equation}\label{eq3}
\mathbb{P}\left(L\leq \beta_j^M\leq U|\hat{M}(y)=\hat{M}(y^{\text{obs}})\right)=1-\alpha
\end{equation}
Indeed, the probability of occurrence of \(y \in \mathbb{R}^n\) satisfying \(\eta^\top y^{\text{obs}}\leq \eta^\top y\) is determined by \(\beta_j^M\). Similarly, under the condition \(\hat{M}(y)=\hat{M}(y^{\text{obs}})\), the conditional probability 
\begin{equation}\label{eq-new001}
\mathbb{P}(\eta^\top y^{\text{obs}}\leq \eta^\top y|\hat{M}(y)=\hat{M}(y^{\text{obs}}))
\end{equation}
holds. The lower limit of \(\beta_j^M\) for which (\ref{eq-new001}) becomes at least \(\alpha/2\) is \(L\).
Similarly, for the conditional probability
\[
\mathbb{P}\left(\eta^\top y^{\text{obs}} \geq \eta^\top y \mid \hat{M}(y) = \hat{M}(y^{\text{obs}})\right)
\]
the upper limit of \(\beta_j^M\) that causes it to be at least \(\alpha/2\) is \(U\).

\begin{figure}
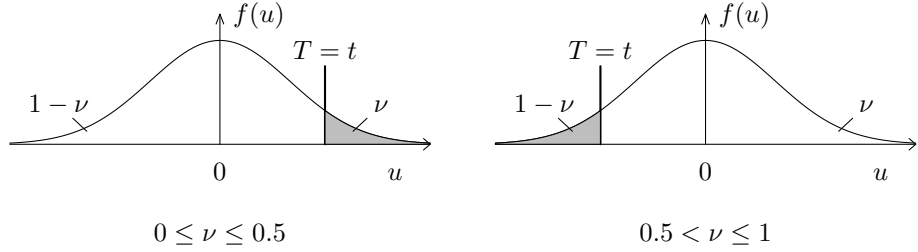

\begin{center}
\begin{tabular}{cc}
\input{fig-new003}&\input{fig-new004}\\
$0\leq \nu\leq 0.5$&$0.5<\nu\leq 1$
\end{tabular}
\end{center}
\caption{\label{fig-new001} 
Regardless of whether the distribution of $t$ is a truncated distribution,
when the null distribution is $f(u)$ and the statistic $T$ is $t$, the value of $\nu=\int_t^\infty f(u)du$ is,
regardless of $0\leq \nu\leq 0.5$, $0.5<\nu\leq 1$, the p-value is $2\min\{\nu,1-\nu\}$.}
\end{figure}

On the other hand, in the case of a two-sided test, the p-value of the statistic \(\hat{\beta}_j^{\hat{M}(y^{\text{obs}})}\) under the null hypothesis \({\beta}_j^{\hat{M}(y^{\text{obs}})}=0\) can be calculated as,
\begin{eqnarray}\label{eq4}
&&2\min\left\{ \mathbb{P}\left(\eta^\top y\leq \eta^\top y^{\text{obs}}|\beta_j^M=0, \hat{M}(y)=\hat{M}(y^{\text{obs}})\right)\right.\ , \nonumber\\
&&\left. \mathbb{P}\left(\eta^\top y^{\text{obs}}\leq \eta^\top y|\beta_j^M=0, \hat{M}(y)=\hat{M}(y^{\text{obs}})\right) \right\}
\end{eqnarray}
In fact, generally, when the null distribution is \(f(t)\) and the statistic \(T\) is \(t\), regardless of whether 
\[
\nu := \int_t^\infty f(u)du
\]
is \(0 \leq \nu \leq 0.5\) or \(0.5 < \nu \leq 1\), \(2\min\{\nu,1-\nu\}\) becomes the \(p\)-value (see Figure \ref{fig-new001}). Therefore, if we set \(T = \eta^\top y^{\text{obs}}\) and \(f\) as the conditional density function of \(\eta^\top y\) under \(\beta_j^M=0\) and \(\hat{M}(y)=\hat{M}(y^{\text{obs}})\), and let 
\[
\nu = \int^\infty_{\eta^\top y^{\text{obs}}}f(t)dt
\]
then (\ref{eq4}) is obtained. Note that \(\hat{M}\) takes the same value even if \(y\) is different as long as \(\eta^\top y\) is the same.

Although it seems repetitive, conducting tests and estimating confidence intervals based on the distribution (truncated distribution) that considers 
\[
\left\{y\in \mathbb{R}^n\ |\ \hat{M}(y)=\hat{M}(y^{\text{obs}})\right\}
\]
as the universal set, rather than
\[
\left\{y\in \mathbb{R}^n\right\}
\]
is referred to as selective inference.

\section{Selective Inference in Lasso}

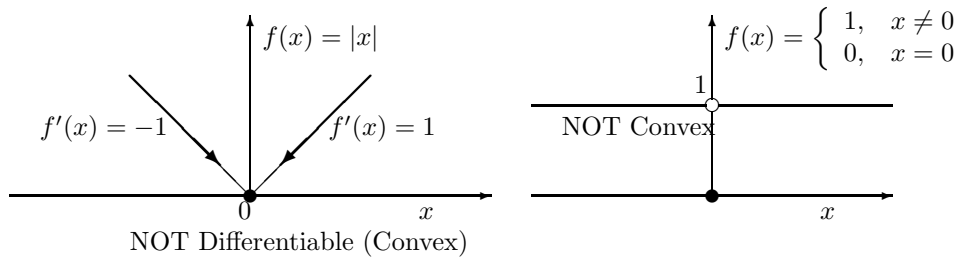
\begin{figure}
\begin{tabular}{cc}
\setlength{\unitlength}{0.8mm}
\begin{picture}(80,60)
\put(20,30){\line(1,-1){20}}
\put(60,30){\line(-1,-1){20}}
\put(0,10){\vector(1,0){80}}
\put(40,10){\vector(0,1){30}}
\put(40,10){\circle*{2}}

\put(42,35){$f(x)=|x|$}

\put(5,20){$f'(x)=-1$}
\put(53,20){$f'(x)=1$}
\put(38,6){0}
\put(68,6){$x$}
\put(20,1){NOT Differentiable (Convex)}

\thicklines
\put(20,30){\vector(1,-1){15}}
\put(60,30){\vector(-1,-1){15}}
\end{picture}&\setlength{\unitlength}{0.8mm}
\begin{picture}(80,60)
\put(0,10){\vector(1,0){60}}
\put(30,10){\line(0,1){14}}
\put(30,26){\vector(0,1){14}}
\put(30,10){\circle*{2}}
\put(30,25){\circle{2}}
\put(27,27){$1$}
\put(5,20){NOT Convex}
\put(32,35){$f(x)=\left\{\begin{array}{ll}1,&x\not=0\\0,&x=0\end{array}\right.$}
\thicklines
\put(0,25){\line(1,0){29}}
\put(31,25){\line(1,0){29}}
\put(48,6){$x$}
\end{picture}
\end{tabular}
\label{fig03}\caption{
Left: $f(x)=|x|$ is convex. However, at the origin, the derivative does not coincide from both sides, and it cannot be differentiated. Understanding that the subdifferential at $x=0$ is $[-1,1]$ might be seen as the derivative value increasing from $-1$ to $1$ at the "dance floor" ($x=0$), since the derivatives for $x<0$ and $x>0$ are $-1$ and $1$ respectively.
Right: Not convex. For example, information criteria like AIC or BIC are not convex as $g(n)$ becomes a function of the number of non-zero parameters. Therefore, searching for the optimal solution is not efficient.}
\end{figure}

Given $X\in {\mathbb R}^{n\times p}$, $y\in {\mathbb R}^n$, for each $\lambda>0$ we consider
\begin{equation}\label{eq71}
\frac{1}{2}\|y-X\beta\|_2^2+\lambda \|\beta\|_1
\end{equation}
to minimize over $\beta\in {\mathbb R}^p$, a problem referred to as the Lasso in linear regression.
Here, $\|\beta\|_1:=\sum_{j=1}^p|\beta_j|$ (typically, normalization is performed for each $j=1,\ldots,p$ to ensure that $\sum_{i=1}^nx_{i,j}^2=1$, before optimizing). Even though we seek to minimize, the inclusion of the absolute value function (non-differentiable at the origin) means that when each column of $X$ is orthogonal, it operates to set $\beta_j=0$ for $j$ where the absolute value of the ordinary least squares estimator $\sum_{i=1}^nx_{i,j}y_i$ is small \citep{sparse_r, sparse_p}. Choosing $j$ such that $\beta_j\not=0$ can be interpreted as performing variable selection. Unlike the case in ordinary least squares, even when $n<p$ and the situation is sparse, a solution can be computed under suitable conditions.

In cases where $n<p$, since $X^\top X$ does not have an inverse matrix, applying ordinary least squares is challenging. Moreover,
letting $g(n)$ be a monotonically increasing function of $n$ that takes positive values, comparing 
\begin{equation}\label{eq281}
\frac{1}{2}\|y-X\beta\|_2^2+g(n)\sum_{j=1}^p I(\beta_j\not=0)
\end{equation}
across models such as $\{\},\{1\},\{2\},\ldots,\{1,\ldots,p\}$, etc., which totals to $2^p$ models, would require immense computations. Here, $I(A)$ takes the value 1 when condition $A$ is true, and 0 otherwise. Both terms in (\ref{eq71}) are convex, and methods exist for efficiently searching for solutions. However, due to the non-convexity of the graph in Figure \ref{fig03} on the right, searching for solutions to (\ref{eq281}) is not straightforward.

\cite{lee} demonstrated that the condition $\hat{M}(y)=\hat{M}(y^{obs})$ in Lasso variable selection can be expressed using some $m\geq 1, A\in {\mathbb R}^{m\times n}, b\in {\mathbb R}^m$ as a set of inequalities $Ay\leq b$.

Generally, for a convex function \(f: {\mathbb R}\rightarrow {\mathbb R}\) at \(a\in {\mathbb R}\), when
\[f(x)\geq f(a) +z(x-a)\ ,\ x\in {\mathbb R}\]
is satisfied, the set of such \(z\in {\mathbb R}\) is called the subdifferential of \(f\) at \(a\).
If \(f(x)\) is differentiable at \(x=a\), then its subdifferential is the set consisting only of \(f'(a)\).
Moreover, if \(f(x)=|x|\) and \(a=0\), then its subdifferential is the interval \([-1,1]\). Indeed,
\[|x|\geq zx \text{ for any } x\in {\mathbb R}\] \[\Longleftrightarrow 
\left\{
\begin{array}{ll}
z\leq 1,&x>0\\
z\geq 1,&x<0
\end{array}
\right.
\Longleftrightarrow
|z|\leq 1\]
is satisfied (see Figure \ref{fig03}).

\begin{figure}
\begin{center}
\begin{tabular}{cc}
\begin{tikzpicture}[x=0.8pt,y=0.8pt]
\definecolor{fillColor}{RGB}{255,255,255}
\path[use as bounding box,fill=fillColor,fill opacity=0.00] (0,0) rectangle (216.81,216.81);
\begin{scope}
\path[clip] ( 49.20, 61.20) rectangle (191.61,167.61);
\definecolor{drawColor}{RGB}{0,0,0}

\path[draw=drawColor,line width= 0.4pt,line join=round,line cap=round] ( 54.47,163.67) --
	( 55.79,161.26) --
	( 57.11,158.87) --
	( 58.43,156.50) --
	( 59.75,154.16) --
	( 61.07,151.85) --
	( 62.39,149.55) --
	( 63.70,147.29) --
	( 65.02,145.04) --
	( 66.34,142.82) --
	( 67.66,140.63) --
	( 68.98,138.46) --
	( 70.30,136.31) --
	( 71.62,134.19) --
	( 72.94,132.09) --
	( 74.25,130.02) --
	( 75.57,127.97) --
	( 76.89,125.94) --
	( 78.21,123.94) --
	( 79.53,121.97) --
	( 80.85,120.01) --
	( 82.17,118.09) --
	( 83.48,116.18) --
	( 84.80,114.30) --
	( 86.12,112.45) --
	( 87.44,110.62) --
	( 88.76,108.81) --
	( 90.08,107.03) --
	( 91.40,105.27) --
	( 92.71,103.53) --
	( 94.03,101.82) --
	( 95.35,100.14) --
	( 96.67, 98.48) --
	( 97.99, 96.84) --
	( 99.31, 95.23) --
	(100.63, 93.64) --
	(101.94, 92.07) --
	(103.26, 90.53) --
	(104.58, 89.02) --
	(105.90, 87.53) --
	(107.22, 86.06) --
	(108.54, 84.62) --
	(109.86, 83.20) --
	(111.17, 81.80) --
	(112.49, 80.43) --
	(113.81, 79.09) --
	(115.13, 77.76) --
	(116.45, 76.47) --
	(117.77, 75.19) --
	(119.09, 73.94) --
	(120.41, 72.72) --
	(121.72, 72.13) --
	(123.04, 71.56) --
	(124.36, 71.01) --
	(125.68, 70.49) --
	(127.00, 69.99) --
	(128.32, 69.52) --
	(129.64, 69.07) --
	(130.95, 68.65) --
	(132.27, 68.25) --
	(133.59, 67.87) --
	(134.91, 67.52) --
	(136.23, 67.19) --
	(137.55, 66.89) --
	(138.87, 66.61) --
	(140.18, 66.35) --
	(141.50, 66.12) --
	(142.82, 65.92) --
	(144.14, 65.74) --
	(145.46, 65.58) --
	(146.78, 65.44) --
	(148.10, 65.34) --
	(149.41, 65.25) --
	(150.73, 65.19) --
	(152.05, 65.15) --
	(153.37, 65.14) --
	(154.69, 65.15) --
	(156.01, 65.19) --
	(157.33, 65.25) --
	(158.64, 65.34) --
	(159.96, 65.44) --
	(161.28, 65.58) --
	(162.60, 65.74) --
	(163.92, 65.92) --
	(165.24, 66.12) --
	(166.56, 66.35) --
	(167.88, 66.61) --
	(169.19, 66.89) --
	(170.51, 67.19) --
	(171.83, 67.52) --
	(173.15, 67.87) --
	(174.47, 68.25) --
	(175.79, 68.65) --
	(177.11, 69.07) --
	(178.42, 69.52) --
	(179.74, 69.99) --
	(181.06, 70.49) --
	(182.38, 71.01) --
	(183.70, 71.56) --
	(185.02, 72.13) --
	(186.34, 72.72);
\end{scope}
\begin{scope}
\path[clip] (  0.00,  0.00) rectangle (216.81,216.81);
\definecolor{drawColor}{RGB}{0,0,0}

\path[draw=drawColor,line width= 0.4pt,line join=round,line cap=round] ( 54.47, 61.20) -- (186.34, 61.20);

\path[draw=drawColor,line width= 0.4pt,line join=round,line cap=round] ( 54.47, 61.20) -- ( 54.47, 55.20);

\path[draw=drawColor,line width= 0.4pt,line join=round,line cap=round] ( 87.44, 61.20) -- ( 87.44, 55.20);

\path[draw=drawColor,line width= 0.4pt,line join=round,line cap=round] (120.41, 61.20) -- (120.41, 55.20);

\path[draw=drawColor,line width= 0.4pt,line join=round,line cap=round] (153.37, 61.20) -- (153.37, 55.20);

\path[draw=drawColor,line width= 0.4pt,line join=round,line cap=round] (186.34, 61.20) -- (186.34, 55.20);

\node[text=drawColor,anchor=base,inner sep=0pt, outer sep=0pt, scale=  1.00] at ( 54.47, 39.60) {-2};

\node[text=drawColor,anchor=base,inner sep=0pt, outer sep=0pt, scale=  1.00] at ( 87.44, 39.60) {-1};

\node[text=drawColor,anchor=base,inner sep=0pt, outer sep=0pt, scale=  1.00] at (120.41, 39.60) {0};

\node[text=drawColor,anchor=base,inner sep=0pt, outer sep=0pt, scale=  1.00] at (153.37, 39.60) {1};

\node[text=drawColor,anchor=base,inner sep=0pt, outer sep=0pt, scale=  1.00] at (186.34, 39.60) {2};

\path[draw=drawColor,line width= 0.4pt,line join=round,line cap=round] ( 49.20, 72.72) -- ( 49.20,163.67);

\path[draw=drawColor,line width= 0.4pt,line join=round,line cap=round] ( 49.20, 72.72) -- ( 43.20, 72.72);

\path[draw=drawColor,line width= 0.4pt,line join=round,line cap=round] ( 49.20, 87.88) -- ( 43.20, 87.88);

\path[draw=drawColor,line width= 0.4pt,line join=round,line cap=round] ( 49.20,103.04) -- ( 43.20,103.04);

\path[draw=drawColor,line width= 0.4pt,line join=round,line cap=round] ( 49.20,118.19) -- ( 43.20,118.19);

\path[draw=drawColor,line width= 0.4pt,line join=round,line cap=round] ( 49.20,133.35) -- ( 43.20,133.35);

\path[draw=drawColor,line width= 0.4pt,line join=round,line cap=round] ( 49.20,148.51) -- ( 43.20,148.51);

\path[draw=drawColor,line width= 0.4pt,line join=round,line cap=round] ( 49.20,163.67) -- ( 43.20,163.67);

\node[text=drawColor,rotate= 90.00,anchor=base,inner sep=0pt, outer sep=0pt, scale=  1.00] at ( 34.80, 72.72) {0};

\node[text=drawColor,rotate= 90.00,anchor=base,inner sep=0pt, outer sep=0pt, scale=  1.00] at ( 34.80, 87.88) {2};

\node[text=drawColor,rotate= 90.00,anchor=base,inner sep=0pt, outer sep=0pt, scale=  1.00] at ( 34.80,103.04) {4};

\node[text=drawColor,rotate= 90.00,anchor=base,inner sep=0pt, outer sep=0pt, scale=  1.00] at ( 34.80,118.19) {6};

\node[text=drawColor,rotate= 90.00,anchor=base,inner sep=0pt, outer sep=0pt, scale=  1.00] at ( 34.80,133.35) {8};

\node[text=drawColor,rotate= 90.00,anchor=base,inner sep=0pt, outer sep=0pt, scale=  1.00] at ( 34.80,148.51) {10};

\node[text=drawColor,rotate= 90.00,anchor=base,inner sep=0pt, outer sep=0pt, scale=  1.00] at ( 34.80,163.67) {12};

\path[draw=drawColor,line width= 0.4pt,line join=round,line cap=round] ( 49.20, 61.20) --
	(191.61, 61.20) --
	(191.61,167.61) --
	( 49.20,167.61) --
	( 49.20, 61.20);
\end{scope}

\begin{scope}
\path[clip] (  0.00,  0.00) rectangle (252.94,252.94);
\definecolor{drawColor}{RGB}{0,0,0}

\node[text=drawColor,anchor=base,inner sep=0pt, outer sep=0pt, scale=  1.20] at (118.47,189.20) {\bfseries $y=x^2-3x+|x|$};

\node[text=drawColor,anchor=base,inner sep=0pt, outer sep=0pt, scale=  1.00] at (120.40, 20.60) {$x$};

\node[text=drawColor,rotate= 90.00,anchor=base,inner sep=0pt, outer sep=0pt, scale=  1.00] at ( 10.80,114.41) {$y$};
\end{scope}

\begin{scope}
\path[clip] (  0.00,  0.00) rectangle (216.81,216.81);
\definecolor{fillColor}{RGB}{0,0,0}

\path[fill=fillColor] (153.37, 65.14) circle (  2.25);
\end{scope}
\end{tikzpicture}&
\begin{tikzpicture}[x=0.8pt,y=0.8pt]
\definecolor{fillColor}{RGB}{255,255,255}
\path[use as bounding box,fill=fillColor,fill opacity=0.00] (0,0) rectangle (216.81,216.81);
\begin{scope}
\path[clip] ( 49.20, 61.20) rectangle (191.61,167.61);
\definecolor{drawColor}{RGB}{0,0,0}

\path[draw=drawColor,line width= 0.4pt,line join=round,line cap=round] ( 54.47,124.26) --
	( 55.79,122.30) --
	( 57.11,120.38) --
	( 58.43,118.49) --
	( 59.75,116.63) --
	( 61.07,114.80) --
	( 62.39,113.00) --
	( 63.70,111.24) --
	( 65.02,109.50) --
	( 66.34,107.80) --
	( 67.66,106.13) --
	( 68.98,104.49) --
	( 70.30,102.88) --
	( 71.62,101.30) --
	( 72.94, 99.76) --
	( 74.25, 98.25) --
	( 75.57, 96.76) --
	( 76.89, 95.31) --
	( 78.21, 93.90) --
	( 79.53, 92.51) --
	( 80.85, 91.15) --
	( 82.17, 89.83) --
	( 83.48, 88.54) --
	( 84.80, 87.27) --
	( 86.12, 86.04) --
	( 87.44, 84.85) --
	( 88.76, 83.68) --
	( 90.08, 82.55) --
	( 91.40, 81.44) --
	( 92.71, 80.37) --
	( 94.03, 79.33) --
	( 95.35, 78.32) --
	( 96.67, 77.34) --
	( 97.99, 76.40) --
	( 99.31, 75.48) --
	(100.63, 74.60) --
	(101.94, 73.75) --
	(103.26, 72.93) --
	(104.58, 72.14) --
	(105.90, 71.38) --
	(107.22, 70.66) --
	(108.54, 69.97) --
	(109.86, 69.30) --
	(111.17, 68.67) --
	(112.49, 68.07) --
	(113.81, 67.51) --
	(115.13, 66.97) --
	(116.45, 66.47) --
	(117.77, 65.99) --
	(119.09, 65.55) --
	(120.41, 65.14) --
	(121.72, 66.34) --
	(123.04, 67.57) --
	(124.36, 68.83) --
	(125.68, 70.12) --
	(127.00, 71.45) --
	(128.32, 72.80) --
	(129.64, 74.19) --
	(130.95, 75.61) --
	(132.27, 77.06) --
	(133.59, 78.54) --
	(134.91, 80.05) --
	(136.23, 81.60) --
	(137.55, 83.18) --
	(138.87, 84.78) --
	(140.18, 86.42) --
	(141.50, 88.09) --
	(142.82, 89.80) --
	(144.14, 91.53) --
	(145.46, 93.30) --
	(146.78, 95.09) --
	(148.10, 96.92) --
	(149.41, 98.78) --
	(150.73,100.67) --
	(152.05,102.60) --
	(153.37,104.55) --
	(154.69,106.54) --
	(156.01,108.56) --
	(157.33,110.61) --
	(158.64,112.69) --
	(159.96,114.80) --
	(161.28,116.94) --
	(162.60,119.12) --
	(163.92,121.33) --
	(165.24,123.56) --
	(166.56,125.83) --
	(167.88,128.14) --
	(169.19,130.47) --
	(170.51,132.83) --
	(171.83,135.23) --
	(173.15,137.66) --
	(174.47,140.12) --
	(175.79,142.61) --
	(177.11,145.13) --
	(178.42,147.68) --
	(179.74,150.27) --
	(181.06,152.89) --
	(182.38,155.53) --
	(183.70,158.21) --
	(185.02,160.93) --
	(186.34,163.67);
\end{scope}
\begin{scope}
\path[clip] (  0.00,  0.00) rectangle (216.81,216.81);
\definecolor{drawColor}{RGB}{0,0,0}

\path[draw=drawColor,line width= 0.4pt,line join=round,line cap=round] ( 54.47, 61.20) -- (186.34, 61.20);

\path[draw=drawColor,line width= 0.4pt,line join=round,line cap=round] ( 54.47, 61.20) -- ( 54.47, 55.20);

\path[draw=drawColor,line width= 0.4pt,line join=round,line cap=round] ( 87.44, 61.20) -- ( 87.44, 55.20);

\path[draw=drawColor,line width= 0.4pt,line join=round,line cap=round] (120.41, 61.20) -- (120.41, 55.20);

\path[draw=drawColor,line width= 0.4pt,line join=round,line cap=round] (153.37, 61.20) -- (153.37, 55.20);

\path[draw=drawColor,line width= 0.4pt,line join=round,line cap=round] (186.34, 61.20) -- (186.34, 55.20);

\node[text=drawColor,anchor=base,inner sep=0pt, outer sep=0pt, scale=  1.00] at ( 54.47, 39.60) {-2};

\node[text=drawColor,anchor=base,inner sep=0pt, outer sep=0pt, scale=  1.00] at ( 87.44, 39.60) {-1};

\node[text=drawColor,anchor=base,inner sep=0pt, outer sep=0pt, scale=  1.00] at (120.41, 39.60) {0};

\node[text=drawColor,anchor=base,inner sep=0pt, outer sep=0pt, scale=  1.00] at (153.37, 39.60) {1};

\node[text=drawColor,anchor=base,inner sep=0pt, outer sep=0pt, scale=  1.00] at (186.34, 39.60) {2};

\path[draw=drawColor,line width= 0.4pt,line join=round,line cap=round] ( 49.20, 65.14) -- ( 49.20,163.67);

\path[draw=drawColor,line width= 0.4pt,line join=round,line cap=round] ( 49.20, 65.14) -- ( 43.20, 65.14);

\path[draw=drawColor,line width= 0.4pt,line join=round,line cap=round] ( 49.20, 84.85) -- ( 43.20, 84.85);

\path[draw=drawColor,line width= 0.4pt,line join=round,line cap=round] ( 49.20,104.55) -- ( 43.20,104.55);

\path[draw=drawColor,line width= 0.4pt,line join=round,line cap=round] ( 49.20,124.26) -- ( 43.20,124.26);

\path[draw=drawColor,line width= 0.4pt,line join=round,line cap=round] ( 49.20,143.96) -- ( 43.20,143.96);

\path[draw=drawColor,line width= 0.4pt,line join=round,line cap=round] ( 49.20,163.67) -- ( 43.20,163.67);

\node[text=drawColor,rotate= 90.00,anchor=base,inner sep=0pt, outer sep=0pt, scale=  1.00] at ( 34.80, 65.14) {0};

\node[text=drawColor,rotate= 90.00,anchor=base,inner sep=0pt, outer sep=0pt, scale=  1.00] at ( 34.80, 84.85) {2};

\node[text=drawColor,rotate= 90.00,anchor=base,inner sep=0pt, outer sep=0pt, scale=  1.00] at ( 34.80,104.55) {4};

\node[text=drawColor,rotate= 90.00,anchor=base,inner sep=0pt, outer sep=0pt, scale=  1.00] at ( 34.80,124.26) {6};

\node[text=drawColor,rotate= 90.00,anchor=base,inner sep=0pt, outer sep=0pt, scale=  1.00] at ( 34.80,143.96) {8};

\node[text=drawColor,rotate= 90.00,anchor=base,inner sep=0pt, outer sep=0pt, scale=  1.00] at ( 34.80,163.67) {10};

\path[draw=drawColor,line width= 0.4pt,line join=round,line cap=round] ( 49.20, 61.20) --
	(191.61, 61.20) --
	(191.61,167.61) --
	( 49.20,167.61) --
	( 49.20, 61.20);
\end{scope}

\begin{scope}
\path[clip] (  0.00,  0.00) rectangle (252.94,252.94);
\definecolor{drawColor}{RGB}{0,0,0}

\node[text=drawColor,anchor=base,inner sep=0pt, outer sep=0pt, scale=  1.20] at (118.47,189.20) {\bfseries $y=x^2+x+2|x|$};

\node[text=drawColor,anchor=base,inner sep=0pt, outer sep=0pt, scale=  1.00] at (120.41, 20.60) {$x$};

\node[text=drawColor,rotate= 90.00,anchor=base,inner sep=0pt, outer sep=0pt, scale=  1.00] at ( 10.80,114.41) {$y$};
\end{scope}

\begin{scope}
\path[clip] (  0.00,  0.00) rectangle (216.81,216.81);
\definecolor{fillColor}{RGB}{0,0,0}

\path[fill=fillColor] (120.41, 65.14) circle (  2.25);
\end{scope}
\end{tikzpicture}
\end{tabular}
\end{center}
\caption{\label{fig04}
For example, in the function on each of the left and right, even though \(x^2-3x\) and \(x^2+x\) are differentiable, because \(|x|\) is not differentiable at \(x=0\), the subdifferential at \(x=0\) becomes the interval \([-4,-2]\) and \([-1,3]\) respectively.
The left does not include 0 in that interval, while the right includes 0. Therefore, the left does not become a minimum at the origin, whereas the right becomes a minimum.}
\end{figure}
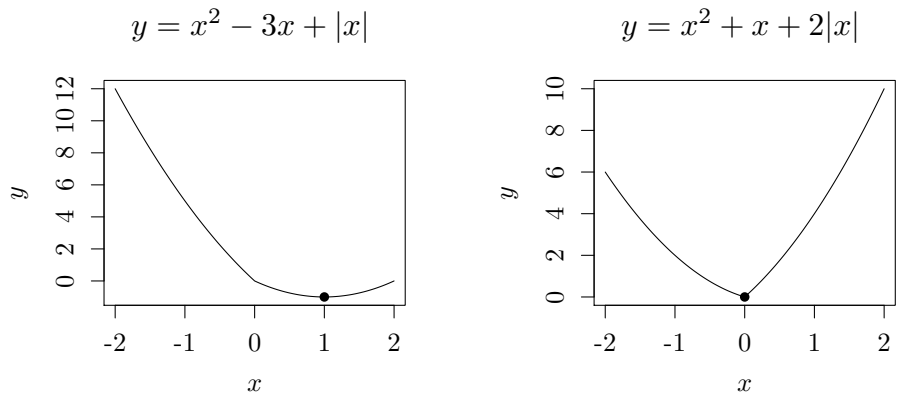

Therefore, when subdifferentiating (\ref{eq71}), we can write
\begin{equation}\label{eq272}
X^\top (X\hat{\beta}-y)+\lambda \hat{s}=0
\end{equation}
where \(\hat{s}=[\hat{s}_1,\ldots,\hat{s}_p]^\top\), and if the solution \(\hat{\beta}_j\) of (\ref{eq272}) is non-zero (active), \(\hat{s}_j={\rm sign}(\hat{\beta}_j)\), and if \(\hat{\beta}_j\) is 0 (inactive), \(\hat{s}_j\in [-1,1]\). An example of how to find the solution of a function containing an absolute value function is shown in Figure \ref{fig04}. Dividing the index set \(\{1,\ldots,p\}\) into
\[
\hat{M}:=\{j|\hat{\beta}_j\not=0\}\ ,\ -\hat{M}:=\{j|\hat{\beta}_j=0\}
\]
assuming there are \(m\) elements in \(\hat{M}\), and the first and last \(m\) columns of \(X\) correspond to the indices of \(\hat{M},-\hat{M}\) respectively. The same applies to \(\hat{\beta},\hat{s}\). Then, representing them as
\[
X=[X_{\hat{M}},X_{-\hat{M}}]\ ,\ 
\hat{\beta}=\left[
\begin{array}{c}
\hat{\beta}_{\hat{M}}\\
\hat{\beta}_{-\hat{M}}
\end{array}
\right]\ ,\ 
\hat{s}=\left[
\begin{array}{c}
\hat{s}_{\hat{M}}\\
\hat{s}_{-\hat{M}}
\end{array}
\right]
\]
we can concretely write (\ref{eq272}) as
\begin{equation}\label{eq11}
X_{\hat{M}}^\top (X_{\hat{M}}\hat{\beta}_{\hat{M}}-y)+\lambda \hat{s}_{\hat{M}}=0
\end{equation}
\begin{equation}\label{eq121}
X_{-\hat{M}}^\top (X_{\hat{M}}\hat{\beta}_{\hat{M}}-y)+\lambda \hat{s}_{-\hat{M}}=0
\end{equation}
\begin{equation}\label{eq13}
{\rm sign}(\hat{\beta}_{\hat{M}})=\hat{s}_{\hat{M}}
\end{equation}
\begin{equation}\label{eq14}
\|\hat{s}_{-\hat{M}}\|_\infty\leq 1
\end{equation}
However, we denote \(\|u\|_\infty\) to represent the maximum absolute value of the components of \(u=[u_1,\ldots,u_p]^\top\). 
Also, if we let \(P^\perp_M:=I-X_M(X^\top_MX_M)^{-1}X_M^\top\), \(X_M^+:=(X_M^\top X_M)^{-1}X_M^\top\), then from (\ref{eq11}) and (\ref{eq121}) we can write
\begin{eqnarray}\label{eq31}
\hat{\beta}_{\hat{M}}=(X_{\hat{M}}^\top X_{\hat{M}})^{-1}(X_{\hat{M}}^\top y-\lambda \hat{s}_{\hat{M}})
\end{eqnarray}
\begin{eqnarray}\label{eq32}
\hat{s}_{-\hat{M}}&=&
-\frac{1}{\lambda}
\{X_{-\hat{M}}^\top (X_{\hat{M}}\hat{\beta}_{\hat{M}}-y)\}\nonumber\\
&=&
-\frac{1}{\lambda}
\{X_{-\hat{M}}^\top (X_{\hat{M}}(X_{\hat{M}}^\top X_{\hat{M}})^{-1}(X_{\hat{M}}^\top y-\lambda \hat{s}_{\hat{M}})-y)\}\nonumber
\\
&=&
X_{-\hat{M}}^\top (X_{\hat{M}}^+)^{\top}\hat{s}_{\hat{M}}+\frac{1}{\lambda}X_{-\hat{M}}^\top P^\perp_{\hat{M}}y
\end{eqnarray}

Therefore, for \((\hat{M},\hat{s}_{\hat{M}})=(M,s)\) to satisfy (\ref{eq13}) and (\ref{eq14}), it is necessary and sufficient from (\ref{eq31}) and (\ref{eq32}) that
\begin{equation}\label{eq310}
\left\{
\begin{array}{ll}
\displaystyle {\rm diag}(s)(X_M^\top X_M)^{-1}(X_M^\top y-\lambda s)>0\\
\displaystyle -1<X_{-{M}}^\top (X_{{M}}^+)^{\top}s+\frac{1}{\lambda}X_{-{M}}^\top P_M^\perp y<1
\end{array}
\right.
\end{equation}

Note that the two inequalities in (\ref{eq310}) indicate that each element of the \(m\)-dimensional and \(p-m\)-dimensional vectors, respectively, fall within that range. Moreover, this can be written as 
$$
\left[
\begin{array}{c}
\frac{1}{\lambda}X_{-M}^\top P_M^\perp\\
-\frac{1}{\lambda}X_{-M}^\top P_M^\perp\\
-{\rm diag}(s)(X_M^+)^\top
\end{array}
\right]
y\leq
\left[
\begin{array}{c}
1_{p-m}-X_{-M}^\top (X_M^\top)^+ s\\
1_{p-m}+X_{-M}^\top (X_M^\top)^+ s\\
-\lambda{\rm diag}(s)(X_M^\top X_M )^{-1}s
\end{array}
\right]\ ,
$$
where \(1_{p-m}\in {\mathbb R}^{p-m}\) is a vector with all components equal to 1, and \({\rm diag}(s)\) is a diagonal matrix with \(s\in \{-1,1\}^m\) as diagonal components. Therefore, 
$$\{\hat{M}=M,\hat{s}_{\hat{M}}=s\}=\{A(M,s)y\leq b(M,s)\}$$
we can observe the existence of \(A(M,s)\in {\mathbb R}^{(2p-m)\times n}\), \(b(M,s)\in {\mathbb R}^{2p-m}\).

Furthermore, although \((M, s)\) is conditioned by the sign \(s\), summing over all signs results in a polyhedral domain (\cite{lee}). 
\begin{equation}\label{eq3105}
\{\hat{M}=M\}=\cup_{s\in \{-1,1\}^{m}}{\left\{\hat{M}=M,\hat{s}_{\hat{M}}=s\right\}}
=\cup_{s\in \{-1,1\}^{m}}{\left\{ A(M,s)y\leq b(M,s) \right\}}
\end{equation}
Here, at the stage of calculating the conditional probability in Lasso, it is often assumed that \(p\) is sufficiently larger than \(n\), and it is presumed that the number of selected variables, \(m\), is also reasonably large. 
Thus, there is a possibility that summing all \(2^m\) elements involves considerable computation. However, conditioning not on \(\{\hat{M}=M\}\) but on \(\{\hat{M}=M,\hat{s}=s\}\) weakens the detection power and widens the confidence interval.

\section{Selective Inference for Distributions Conditioned on Polyhedra}

Let \(\eta := X(X^\top X)^{-1}e_j\). The parameter \(\beta_j\) of linear regression is often estimated by the least squares method as \(\hat{\beta}_j = \eta^\top y\). In the following, considering \(\eta \in {\mathbb R}^n\), we will examine the conditional distribution of \(\eta^\top y\) under \(\{Ay \leq b\}\). First, let \(\Sigma = \sigma^2 I_n \in {\mathbb R}^{n\times n}\), \(c := \Sigma\eta(\eta^\top \Sigma \eta)^{-1} = \eta/\|\eta\|^2 \in {\mathbb R}^{n}\), and \(z := (I_n - c\eta^\top)y \in {\mathbb R}^n\). The covariance matrix \({\mathbb R}^{1\times p}\) of \(\eta^\top y\) and \(z\), which follow a normal distribution, is
\[ {\rm Cov}[\eta^\top y, z] = {\mathbb E}[\eta^\top (y-X\beta)(y-X\beta)^\top (I_n - c\eta^\top)^\top] = \eta^\top \sigma^2\left(I_n - \frac{\eta\eta^\top}{\|\eta\|^2}\right) = \eta^\top \sigma^2 - \eta^\top \sigma^2 = 0, \]
revealing that \(\eta^\top y\) and \(z\) are independent. Moreover, representing the \(j\)th component of a vector by \((\cdot)_j\), we have
\begin{eqnarray*}
Ay \leq b &\Longleftrightarrow& A(c\eta^\top y + z) \leq b \Longleftrightarrow (Ac)_j(\eta^\top y) \leq b_j - (Az)_j, j = 1,\ldots, p \\
& \Longleftrightarrow & \left\{ \begin{array}{ll} \displaystyle \eta^\top y \leq \frac{b_j - (Az)_j}{(Ac)_j}, & \displaystyle (Ac)_j > 0 \\ \displaystyle \eta^\top y \geq \frac{b_j - (Az)_j}{(Ac)_j}, & \displaystyle (Ac)_j < 0 \\ \displaystyle 0 \leq b_j - (Az)_j, & \displaystyle (Ac)_j = 0 \end{array} \right. \\
& \Longleftrightarrow & \left\{ \begin{array}{l} \displaystyle \eta^\top y \leq \min_{j: (Ac)_j > 0}\frac{b_j - (Az)_j}{(Ac)_j} \\ \displaystyle \eta^\top y \geq \max_{j: (Ac)_j < 0}\frac{b_j - (Az)_j}{(Ac)_j} \\ \displaystyle 0 \leq \min_{j: (Ac)_j = 0}\{b_j - (Az)_j\} \end{array} \right.
\end{eqnarray*}
thus,
\[ \{Ay \leq b\} = \{\nu^-(z) \leq \eta^\top y \leq \nu^+(z), \nu^0(z) \geq 0\} \]
is permissible to write. Where,
\begin{equation}\label{eq91}
\left\{ \begin{array}{l} \nu^-(z) := \displaystyle \max_{j: (Ac)_j < 0}\frac{b_j - (Az)_j}{(Ac)_j} \\ \nu^+(z) := \displaystyle \min_{j: (Ac)_j > 0}\frac{b_j - (Az)_j}{(Ac)_j} \\ \nu^0(z) := \displaystyle \min_{j: (Ac)_j = 0}\{b_j - (Az)_j\} \end{array} \right.
\end{equation}
varies independently of \(\eta^\top y\).
Additionally, since $\eta^\top y$ and $z$ are independent, for $z_0\in {\mathbb R}^n$ such that $\nu^0(z_0)\geq 0$, we have
\begin{equation}\label{eq415}
\eta^\top y | \{Ay\leq b, z=z_0\}=\eta^\top y |  \{\nu^{-}(z_0)\leq \eta^\top y \leq \nu^+(z_0)\}
\end{equation}
where the right side of (\ref{eq415}) becomes a truncated normal distribution in the interval $[\nu^-(z_0),\nu^+(z_0)]$ for $N(\beta_j,\sigma^2\|\eta\|^2)$. Henceforth, we will use the notation
\begin{equation}\label{eq1721}
F^{[a,b]}_{\mu,\sigma^2}(x):=\frac{\Psi((x-\mu)/\sigma)-\Psi((a-\mu)/\sigma)}{\Psi((b-\mu)/\sigma)-\Psi((a-\mu)/\sigma)}
\end{equation}
where $\Psi$ represents the cumulative probability of the standard normal distribution. Equation (\ref{eq1721}) becomes the distribution function of the truncated normal distribution. Generally, the value of the distribution function $F_X(x)$ of a random variable $X$ uniformly distributes in $[0,1]$ when $x$ varies according to $F_X$. Therefore, when $\eta^\top y$ ranges within $[\nu^{-}(z_0), \nu^+(z_0)]$ under $z=z_0$, the value of its distribution function
\begin{equation}\label{eq19}
F^{[\nu^{-}(z), \nu^+(z)]}_{\beta_j,\sigma^2\|\eta\|^2}\left(\eta^\top y|Ay\leq b,z=z_0\right)
\end{equation}
uniformly distributes in $[0,1]$. Here, since $\eta^\top y$ and $z$ are independent, (\ref{eq19}) holds whether $z=z_0$ or not\footnote{In \cite{lee}, the probability concerning $z=z_0$ is multiplied and integrated to marginalize it, thus excluding the impact of $z=z_0$.}.

In practice, taking into account (\ref{eq3105}), it becomes necessary to marginalize over $s \in \{-1,1\}^m$. As a result, we need to consider the logical OR of $2^m$ conditions. When defined as
\[ F_\mu(x) := F^{\cup_s [\nu_s^{-}(z), \nu_s^+(z)]}_{\mu,\sigma^2\|\eta\|^2}(x) |\cup_s\{A_s\leq b_s\} \]
$F_{\beta_j}(\eta^\top y^{obs})$ then becomes uniformly distributed in $[0,1]$. Note that $\nu_s^-(\cdot)$, $\nu_s^+(\cdot)$, $A_s$, and $b_s$ are, respectively, the $\nu^-(\cdot)$, $\nu^+(\cdot)$, $A$, and $b$ for $s \in \{-1,1\}^m$. Thus, the range of $\beta_j$ for which
\[ \frac{\alpha}{2}\leq F_{\beta_j}(\eta^\top y^{obs})\leq 1-\frac{\alpha}{2} \]
holds is the confidence interval at a confidence level of $1-\alpha$. In this context, note the validity of the following lemma.
\begin{hodai}
When $\nu^{-}(z^{obs})\leq x\leq \nu^+(z^{obs})$ is fixed, $F_\mu(x)$ monotonically decreases with respect to $\mu \in \mathbb{R}$.
\end{hodai}
(Proof can be found in the appendix.)

In other words, similar to Section 2,
we should determine \(L, U\) such that 
\[ F_L(\eta^\top y^{obs})=1-\frac{\alpha}{2}, \]
\[ F_U(\eta^\top y^{obs})=\frac{\alpha}{2}. \]
Moreover, the p-value when \(\beta_0\) is fixed is given by
\[ 2\min\{F_0(\eta^\top y^{obs}),1-F_0(\eta^\top y^{obs})\}. \]

However, as seen in (\ref{eq3105}),
computing the logical OR of \(2^m\) conditions is not realistic, especially when \(p\) is large.
Furthermore, setting the condition with a particular single sign can reduce the detection power. Thus,
\cite{takeuchi} proposed a method for constructing the truncated distribution conditioned by a certain parameter value.
First, for each \(j \in \hat{M}(y^{obs})\),
let \(c = \Sigma \eta(\eta^\top \Sigma\eta)^{-1}\), \(z = (I_n-c\eta^\top)y^{obs}\), and \(y = z + cu\).
Note that \(u = \eta^\top y \in \mathbb{R}\) is predetermined to move within the range \([u_{\min}, u_{\max}]\).
Then, specify the range of \(u \in [u_{\min}, u_{\max}]\) such that \(\hat{M}(y) = \hat{M}(y^{obs})\) and determine the truncated distribution.
The reason for seeking \(u\) in the range of the line \(z + cu\) with this method is because \(\eta^\top y, z\) are independent, and there is no need to explore with \(y = z' + cu\) for \(z\) different from \(z'\).

\section{Selective Inference in Forward Stepwise}

In this section, we introduce a method of sequentially selecting variables in linear regression (Forward Stepwise, FS). Although it does not optimize the squared error, it facilitates theoretical analysis. Finally, we determine the polyhedron when applying FS.

Let's define, when \(M_k \subseteq \{1,\ldots,p\}\), \(k=1,2,\ldots,p\),
\[ P_k := X_{M_k}(X^\top_{M_k} X_{M_k})^{-1}X^\top_{M_k}, \]
\[ P_k^\perp := I_n - X_{M_k}(X^\top_{M_k} X_{M_k})^{-1}X^\top_{M_k}. \]
In the following, we utilize that \(P_k^2 = P_k\), \((P_k^\perp)^2 = P_k^\perp\), and \(P_k^\perp P_k = O\).

Departing from Lasso and applying the least squares method,
in case of performing variable selection that minimizes information criteria such as AIC or BIC, 
if there are \(p\) variables in total, there is a method that compares all \(2^p\) models \(M \subseteq \{1,\ldots,p\}\). 
However, this would be computationally intensive. FS, instead, selects a variable \(j_1\) that minimizes the squared error, 
then selects \(j_2\) that minimizes the squared error for the variable set \(\{j_1, j_2\}\), and so on, 
adding one variable \(j_k\) at each \(k\) without backtracking, and halts if the value of the information criterion increases in a naive search method. FS, while not always selecting the optimal model, is often used in analyzing the performance of information criteria.

The procedure of FS (Forward Stepwise) can be described as follows. Initially, define \(M_0 = \{\}\) (empty set), \(r_0 = y\), and \(P_0^\perp = I_n\). Generally, for \(k \geq 1\) and \(j \not\in M_{k-1}\), define the coefficient 
\begin{equation}\label{eq912}
\hat{\beta}_j := \frac{(P^\perp_{k-1} X_{j})^\top r_{k-1}}{\|P^\perp_{k-1}X_{j}\|_2^2},
\end{equation}
and the residual 
\begin{equation}\label{eq913}
q_{j,k-1} := r_{k-1} - P^\perp_{k-1} X_{j} \hat{\beta}_j = \left\{I_n - \frac{P^\perp_{k-1}X_{j}(P^\perp_{k-1}X_{j})^\top}{\|P^\perp_{k-1} X_{j}\|_2^2}\right\}r_{k-1},
\end{equation}
minimize the squared norm \(\|q_{j,k-1}\|_2^2\) to determine \(j_k\), and set \(M_k := M_{k-1}\cup \{j_k\}\). Furthermore, 
\begin{equation}\label{eq601}
r_k := q_{j_k,k-1},
\end{equation}
and increment \(k := k + 1\) for the next cycle.

To provide a supplementary note since the notation using \(P_k, P_k^\perp\) in FS is distinctive: At first, select \(j\) that minimizes the residual \(q_{j,0}\), denoting that \(j\) as \(j_1\), and determine the residual \(r_1\). Next, coordinate-transform \(X\) to \(P_{1}^\perp X\). Then, from among \(P_{1}^\perp X_j\), \(j \neq j_1\), choose \(j\) that minimizes the squared norm of the residual \(q_{j,1}\) as \(j_2\), and find the residual \(r_2\). This process is repeated. Up to steps \(k = 1,2,\ldots,p\), the residual \(P_{k}^\perp y\) when \(X\) is set as \(X_{M_k}\) is obtained. That is, 
\begin{equation}\label{eq221}
r_{k-1} = P^\perp_{k-1}y
\end{equation}
is valid for \(k=1,2,\ldots,p\).

In the subsequent analysis, a polyhedron in the case of Forward Selection (FS) is sought. Initially,
\begin{equation}\label{eq:q_norm}
\|q_{j,k-1}\|_2^2 = \left\| \left\{
I - \frac{P^\perp_{k-1}X_{j}(P^\perp_{k-1}X_{j})^\top}{\|P^\perp_{k-1}X_{j} \|_2^2}
\right\}r_{k-1}\right\|_2^2 = 
r_{k-1}^\top r_{k-1} - 
\frac{r_{k-1}^\top  P^\perp_{k-1}X_{j} (P^\perp_{k-1}X_{j})^\top r_{k-1}}{\|P^\perp_{k-1}X_{j} \|_2^2}
\end{equation}
is established. Stemming from (\ref{eq:q_norm}), the sign of $X_{j_k}^\top P^\perp_{k-1}r_{k-1}$ coincides with the sign of $\hat{\beta}_{j_k}$, denoted as $s_k$. Therefore,
\begin{equation}
s_k = {\rm sign}\left(X_{j_k}^\top P^\perp_{k-1}r_{k-1}\right)
\end{equation}
holds true.
From (\ref{eq221}), for each \(j \not\in M_{k-1}\), we have the following equivalencies,
\begin{eqnarray*}
&& \|q_{j_k,k-1}\|_2^2 \leq \|q_{j,k-1}\|_2^2 \\
&\Longleftrightarrow& \left|\frac{X_{j_k}^\top P^\perp_{k-1}r_{k-1}}{\|X_{j_k}^\top P^\perp_{k-1}\|_2}\right| \geq \left| \frac{X_{j}^\top P^\perp_{k-1}r_{k-1}}{\|X_{j}^\top P^\perp_{k-1}\|_2}\right| \\
&\Longleftrightarrow& \left|\frac{X_{j_k}^\top (P^\perp_{k-1})^2y}{\|X_{j_k}^\top P^\perp_{k-1}\|_2}\right| \geq \left| \frac{X_{j}^\top (P^\perp_{k-1})^2y}{\|X_{j}^\top P^\perp_{k-1}\|_2}\right| \\
&\Longleftrightarrow& s_k\frac{X_{j_k}^\top P^\perp_{k-1}}{\|X_{j_k}^\top P^\perp_{k-1}\|_2}y \geq \frac{X_{j}^\top P^\perp_{k-1}}{\|X_{j}^\top P^\perp_{k-1}\|_2}y\ ,\ 
s_k\frac{X_{j_k}^\top P^\perp_{k-1}}{\|X_{j_k}^\top P^\perp_{k-1}\|_2}y \geq - \frac{X_{j}^\top P^\perp_{k-1}}{\|X_{j}^\top P^\perp_{k-1}\|_2}y
\end{eqnarray*}
will hold true.
Therefore, for \(j \not\in M_{k}\),
\[
\left(\pm \frac{X_{j}^\top P^\perp_{k-1}}{\|X_{j}^\top P_{k-1}^\perp \|_2} - s_k\frac{X_{j_k}^\top P^\perp_{k-1}}{\|X_{j_k}^\top P_{k-1}^\perp \|_2}\right)y \leq 0
\]
Consequently, employing some \(A_k \in \mathbb{R}^{2(p-k)\times n}\) and \(0 = b_k \in \mathbb{R}^{2(p-k)}\), a polyhedron \(A_ky \leq b_k\) can be constructed \cite{ryan_exact}. Moreover, if this procedure is executed up to the \(m\)th step, summing the rows of this inequality on both sides for \(k=1,\ldots,m\) and from \(\sum_{k=1}^m 2(p-k) = 2pm-m^2-m\), we can derive \(Ay \leq b\) where \(A \in \mathbb{R}^{(2pm-m^2-m)\times n}, b \in \mathbb{R}^{2pm-m^2-m}\).

\begin{figure}
\begin{center}
\input{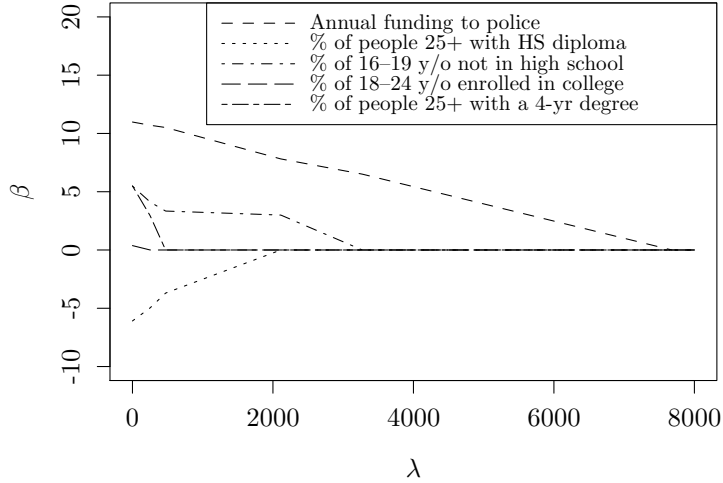}
\end{center}
\caption{\label{fig1-7}
U.S. crime data is stored in the text file \texttt{crime.txt}, and the crime rate per one million people is used as the dependent variable. Lasso and LARS were performed to select explanatory variables from the following. As a result, the same solution path was observed for each \(\lambda\).
}
\end{figure}

\cite{loftus, loftus2} generalized the aforementioned FS concept and proposed a method to divide \(p\) variables into \(G\) groups. Furthermore, extending \cite{loftus}, \cite{yang} has considered a method applicable to Group Lasso.

\section{Selective Inference in LARS}
\begin{figure}
\begin{center}
\setlength{\unitlength}{0.8mm}
\begin{picture}(130,60)
\put(50,10){\circle*{2}}
\put(10,10){\vector(1,0){50}}
\put(10,10){\vector(0,1){30}}
\put(70,10){\vector(1,0){60}}
\put(70,10){\vector(0,1){30}}
\put(48,3){$\lambda_1$}

\put(110,10){\circle*{2}}
\put(110,20){\circle*{2}}
\put(90,10){\circle*{2}}
\put(90,20){\circle*{2}}
\put(90,33){\circle*{2}}
\put(88,3){$\lambda_{k+1}$}
\put(108,3){$\lambda_{k}$}

\put(77,20){\large $\beta_{k}$}
\put(113,20){\large $\beta_{k-1}$}
\put(40,15){\large $\beta_{1}$}

\put(13,40){Coefficients}
\put(73,40){Coefficients}
\put(60,13){$\lambda$}
\put(130,13){$\lambda$}

\thicklines
\put(50,10){\line(-1,1){5}}
\put(90,10){\line(-3,1){5}}
\put(90,20){\line(-2,1){5}}
\put(90,33){\line(-3,1){5}}
\put(110,10){\line(-2,1){20}}
\put(110,20){\line(-3,2){20}}
\put(115,15){\line(-1,1){5}}

\end{picture}
\end{center}
\caption{LARS: 
As $\lambda$ decreases from $\infty$, it reaches $\lambda_1=\max_j|X_j^\top y|$. 
For $j_1$ such that $\lambda_1=|X_j^\top y|$, continue to reduce $\lambda$ while maintaining $\beta_j=0$ for $j \neq j_1$ and only increasing the absolute value of $\beta_1$ (left).
Then, at $\lambda=\lambda_k$, add $j_k \not\in M_{k-1}$ such that $|X_j^\top r(\lambda_k)|=\lambda_k$ to obtain $M_k$, 
and further, with $\beta_j=0$ for $j \not\in M_k$, only increase the absolute values of $\beta_j$ for $j \in M_k$.
}
\label{fig5-115}
\end{figure}
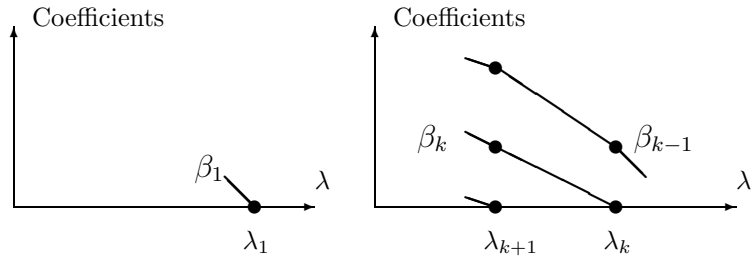

The LARS (Least Angle Regression) to be discussed in this section is a variable selection method similar yet distinct to Lasso. Both FS (Forward Selection) and LARS are considered theoretically amenable to analysis. In the following, we define the procedure of LARS and determine the polyhedron of the truncation region. This polyhedron will also be necessary when considering sequential selective inference in Section 8.

In Lasso, with $\lambda \geq 0$ taken on the horizontal axis\footnote{There are packages that allow displaying $\log \lambda$ or $\|\beta\|_1$ on the horizontal axis}, and the solutions $\hat{\beta}_j(\lambda)$, $j=1,\ldots,p$ for each $\lambda$ obtained from (\ref{eq71}) taken on the vertical axis, the $p$ curves formed are referred to as the solution path. For example, Figure \ref{fig1-7} applies sparse estimation of linear regression to U.S. crime data\footnote{https://web.stanford.edu/ $\tilde{}$ hastie/StatLearnSparsity/data.html}. LARS, as FS does, proceeds through a greedy search variable selection method and, sharing similar properties with Lasso but being theoretically more manageable, is often used when analytically approximating Lasso. While it requires $O(p^3)$ computational time, Lasso has computational efficiency merits.

In LARS, firstly, a piecewise linear function $\beta: [0,\infty) \rightarrow \mathbb{R}^p$ is constructed as follows (see Figure \ref{fig5-115}). Assuming $k \geq 1$, given $r_1 = y, r_2, \ldots, r_{k-1} \in \mathbb{R}^n$, $\lambda_1 := \max_j |X_j^\top y| > \lambda_2 > \ldots > \lambda_{k-1} \geq 0$, $\beta_1 = 0, \beta_2, \ldots, \beta_{k-1} \in \mathbb{R}^p$, $M_{k-1} = \{j_1,\ldots,j_{k-1}\} \subseteq \{1,\ldots,p\}$ (where $j_1$ maximizes $\max_j |X_j^\top y|$), for each $k = 2, 3, \ldots$, the following is performed. Note that $P_{k-1} = X_{M_{k-1}}(X_{M_{k-1}}^\top X_{M_{k-1}})^{-1}X_{M_{k-1}}^\top$.
\begin{enumerate}
\item 
Define the following functions in the range $\lambda\leq \lambda_{k-1}$:
\begin{align}
\beta(\lambda)&=\beta_{k-1}+(1-\frac{\lambda}{\lambda_{k-1}})
\left[ 
\begin{array}{c}
X_{M_{k-1}}^+r_{k-1}\\
0
\end{array}
\right]
\label{eq4-903}
\end{align}
\begin{equation}\label{eq4-904}
r(\lambda)=y-X\beta(\lambda)
\end{equation}
\item Include $j_k \not\in M_{k-1}$ in $M_{k-1}$ to make $M_k$, such that the absolute value of $X_j^\top r(\lambda_k)$ attains the maximum value $\lambda_k$ ($\leq \lambda_{k-1}$).
\item Extend the range of the functions $\beta(\lambda)$, $r(\lambda)$ from $\lambda_{k-1}\leq \lambda$ to $\lambda_k\leq \lambda$, and let $\beta_k:=\beta(\lambda_k)$, $r_{k}:=r(\lambda_{k})$.
\end{enumerate}
Note that in (\ref{eq4-903}), we have denoted the case where the first $k-1$ components are non-zero to prevent the notation from becoming cumbersome (actually, we have not performed any component swapping, etc.). According to this procedure, for $\lambda\leq \lambda_k$ and each $j\in M_k$, 
\begin{equation}\label{eq10}
X_j^\top r(\lambda)=\pm \lambda
\end{equation}
holds. Indeed, from (\ref{eq4-903}) (\ref{eq4-904}),
\begin{eqnarray*}
r(\lambda)&=&y-X\beta(\lambda)=y-X\left\{\beta_{k}+\left(1-\frac{\lambda}{\lambda_{k}}\right)
\left[ 
\begin{array}{c}
X_{M_{k}}^+r_{k}\\
0
\end{array}
\right]
\right\}\\
&=&r_{k}-\left(1-\frac{\lambda}{\lambda_{k}}\right)P_{k}r_{k}=P^\perp_{k}r_{k}+\frac{\lambda}{\lambda_{k}}P_{k}r_{k}
\end{eqnarray*}
is established. Note that from the second step of the algorithm above, $X_j^\top r_{k} =\pm \lambda_{k}$, and $X_j^\top P_{k}=(P_{k}X_j)^\top=X_j^\top$ holds for $j\in M_k$. This means that
$$X_j^\top r(\lambda) =X_j^\top \left(P_{k}^\perp r_{k}+\frac{\lambda}{\lambda_{k}}P_{k}r_{k}\right)
=0+\frac{\lambda}{\lambda_{k}}X_j^\top r_{k}=\pm \lambda$$
In other words, as the value of $\lambda$ decreases and once $j\in M$ becomes active, it continues to satisfy (\ref{eq10}) until $\lambda$ becomes 0, and does not become inactive thereafter.

In selective inference within LARS, variables are sequentially selected, similar to FS. However, we consider the truncation domain and truncation distribution when deciding to stop the selection at any given stage.

The polytope of LARS is constructed as follows. Firstly, by substituting $\lambda = \lambda_k$ into equation (\ref{eq4-903}), we establish

\begin{equation}\label{eq322}
r_k = y - X\beta(\lambda_k) = y - X\left\{\beta_{k-1} + \left(1-\frac{\lambda_k}{\lambda_{k-1}}\right)\right\}\left[ 
\begin{array}{c}
X_{M_{k-1}}^+r_{k-1}\\
0
\end{array}
\right] = r_{k-1} - \left(1-\frac{\lambda_k}{\lambda_{k-1}}\right)P_{k-1}r_{k-1}
\end{equation}

Also, from equation (\ref{eq10}), $r_{k-1}$, $\lambda_{k-1}$ must satisfy $X_{h}^\top r_{k-1}=s_{h}\lambda_{k-1}$, $h\in M_{k-1}$ \cite{sparse_r, sparse_p}, namely, defining $s_{M_{k-1}}:=[s_1,\ldots,s_{k-1}]^\top$, 

\begin{equation}\label{eq321}
X_{M_{k-1}}^\top r_{k-1}=\lambda_{k-1}s_{M_{k-1}}
\end{equation}
must hold. And, pay attention to the following lemma.
\begin{hodai}\label{hodai-new}
\begin{equation}\label{eq12}
X_{j}^\top r_k=s_k\lambda_k \Longleftrightarrow \frac{X_{j}^\top P_{k-1}^\perp}{s_k-X_j^\top (X_{M_{k-1}}^+)^\top s_{M_{k-1}}}y=\lambda_k
\end{equation}
\end{hodai}
(Refer to the appendix for the proof.)

Then, 
\begin{equation}\label{eq107}
c_k(j,s):=\frac{P_{k-1}^\perp X_{j}}{s-X_j^\top (X_{M_{k-1}}^+)^\top s_{M_{k-1}}}
\end{equation}
let's define as above. From (\ref{eq12}), we have
\[ X_{j_k}^\top r_k=s_k\lambda_k \Longleftrightarrow
c_k(j_k,s_k)^\top y=\lambda_k \]

Also, 
\begin{equation}\label{eq41}
c_k(j,s)^\top y\leq \lambda_{k-1}\ ,\ (j,s)\in C_k
\end{equation}
the set of $(j,s)\not\in M_{k-1}\times \{-1,1\}$ for which holds, we denote by $C_{k}$ (the competitors), and the set of $(j,s)\not\in M_{k-1}\times \{-1,1\}$ for which (\ref{eq41}) does not hold, we write as $D_{k}$. It holds that $C_k\cup D_k=\overline{M}_k\times \{-1,1\}$, and
\begin{equation}\label{eq42}
c_k(j,s)^\top y\geq  \lambda_{k-1}\ ,\ (j,s)\in D_k
\end{equation}
\begin{equation}\label{eq43}
c_k(j_k,s_k)^\top y\geq c_k(j,s)^\top y\ ,\ (j,s)\in C_{k}\backslash (j_k,s_k)
\end{equation}
are satisfied \cite{ryan_exact}.
Furthermore, because $\lambda\geq 0$, 
\begin{equation}\label{eq44}
c_k(j_k,s_k)^\top y\geq 0
\end{equation}
holds. That is to say, because the value of $\lambda$ is being lowered from $\lambda_{k-1}$, if $c_k(j,s)^\top y \geq \lambda_{k-1}$, it is excluded from the objects to be compared for the maximum value (becoming an element of $C_k$). Therefore,
$c_k(j_k,s_k)^\top y \leq \lambda_{k-1}$.

At each $k$, the polyhedron is formed by at most $|C_k| + 2(p-k+1)$ inequalities of these four types and the rows of $A,b$ are formed by adding them for $k = 1, \ldots, p$.
However, this polyhedron is the one under the condition of active sets $M_1, \ldots, M_k$, the corresponding signs $s_{M_{1}}, \ldots, s_{M_{k}}$, and competitors $C_1, \ldots, C_k$.
Of course, if the logical sum is taken for possible $\{C_i\}$, the polyhedron related to $\{M_i\}$ and $\{s_{M_i}\}$ is formed, and further, if the logical sum is taken for possible signs $\{s_{M_i}\}$, the polyhedron related to $\{M_i\}$ is formed.

On the other hand, in the case of Lasso, the condition for $(j_k,s_k)$ to become active from non-active is the same.
However, even if it becomes active once, it can become non-active \cite{ryan_unique}.

\section{Significance Test for Lasso}

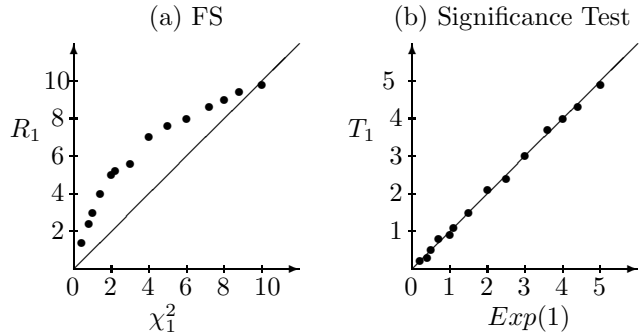
\begin{figure}
\begin{center}
\setlength{\unitlength}{0.5mm}
\begin{picture}(160,80)
\put(30,75){(a) FS}
\put(8,3){0}
\put(18,3){2}
\put(28,3){4}
\put(38,3){6}
\put(48,3){8}
\put(58,3){10}
\put(4,18){2}
\put(4,28){4}
\put(4,38){6}
\put(4,48){8}
\put(2,58){10}
\put(11,20){\line(-1,0){2}}
\put(11,30){\line(-1,0){2}}
\put(11,40){\line(-1,0){2}}
\put(11,50){\line(-1,0){2}}
\put(11,60){\line(-1,0){2}}
\put(-7,45){$R_1$}
\put(30,-5){$\chi^2_1$}
\put(20,11){\line(0,-1){2}}
\put(30,11){\line(0,-1){2}}
\put(40,11){\line(0,-1){2}}
\put(50,11){\line(0,-1){2}}
\put(60,11){\line(0,-1){2}}
\put(10,10){\vector(1,0){60}}
\put(10,10){\vector(0,1){60}}
\put(10,10){\line(1,1){60}}
\put(12,17){\circle*{2}}
\put(14,22){\circle*{2}}
\put(15,25){\circle*{2}}
\put(17,30){\circle*{2}}
\put(20,35){\circle*{2}}
\put(21,36){\circle*{2}}
\put(25,38){\circle*{2}}
\put(30,45){\circle*{2}}
\put(35,48){\circle*{2}}
\put(40,50){\circle*{2}}
\put(46,53){\circle*{2}}
\put(50,55){\circle*{2}}
\put(54,57){\circle*{2}}
\put(60,59){\circle*{2}}

\put(95,75){(b) Significance Test}

\put(98,3){0}
\put(108,3){1}
\put(118,3){2}
\put(128,3){3}
\put(138,3){4}
\put(148,3){5}
\put(94,18){1}
\put(94,28){2}
\put(94,38){3}
\put(94,48){4}
\put(92,58){5}
\put(101,20){\line(-1,0){2}}
\put(101,30){\line(-1,0){2}}
\put(101,40){\line(-1,0){2}}
\put(101,50){\line(-1,0){2}}
\put(101,60){\line(-1,0){2}}
\put(83,45){$T_1$}
\put(120,-5){$Exp(1)$}
\put(110,11){\line(0,-1){2}}
\put(120,11){\line(0,-1){2}}
\put(130,11){\line(0,-1){2}}
\put(140,11){\line(0,-1){2}}
\put(150,11){\line(0,-1){2}}
\put(100,10){\vector(1,0){60}}
\put(100,10){\vector(0,1){60}}
\put(100,10){\line(1,1){60}}
\put(102,12){\circle*{2}}
\put(104,13){\circle*{2}}
\put(105,15){\circle*{2}}
\put(107,18){\circle*{2}}
\put(110,19){\circle*{2}}
\put(111,21){\circle*{2}}
\put(115,25){\circle*{2}}
\put(120,31){\circle*{2}}
\put(125,34){\circle*{2}}
\put(130,40){\circle*{2}}
\put(136,47){\circle*{2}}
\put(140,50){\circle*{2}}
\put(144,53){\circle*{2}}
\put(150,59){\circle*{2}}

\end{picture} 
\end{center}
\caption{\label{fig09}
(a) Values of $R_1$, the FS of the first selected variable among $p$ variables, and (b) values of $T_1$ for the first variable selected in the Significance Test for Lasso, along with the $Q-Q$ plot of the $\chi^2$ values, schematically representing the results of the simulation. The statistic $R_1$ is distributed with larger values than $\chi^2_1$, while the statistic $T_1$ follows a distribution similar to $Exp(1)$.
}
\end{figure}

In this section, we consider sequential statistical tests that do not depend on selective inference using the polyhedra dealt with so far. When $X^\top X = I_p$, the statistic $T_k$, which signifies whether the $k$-th variable is significant or not, can be calculated only from $\lambda_k$, $\lambda_{k+1}$, and $\sigma^2$. This test is shown to have performance asymptotically equivalent to the Spacing Test using a polyhedron, as discussed in Section 8.

In AIC and BIC, for a variable set $M \subseteq \{1, \ldots, p\}$ and $j \not\in M$, it is common to test whether it is significant to add variable $j$ by comparing the difference in the $RSS$ values of $M$ and $M \cup \{j\}$ with $2$ or $\log n$. However, making such a decision based on the value
$$
R_k := \frac{RSS_{M_{k-1}} - RSS_{M_{k-1} \cup \{j_k\}}}{\sigma^2}
$$
normalized by the difference $r_k$ (the value of RSS) defined in (\ref{eq601}) when sequentially selecting variables in FS and adding variable $j$ to the first selected $k-1$ variables has problems. Figure \ref{fig09}(a) is a $Q-Q$ plot with a $\chi^2$ distribution of degree 1, measuring the value of $R_1$ of the first selected variable by performing FS in linear regression where all slopes of $p$ variables are 0, for $n$ times. Since it selects one variable that reduces the RSS the most among $p$ variables, $R_1$ does not follow $\chi^2_1$, even though that variable is not significant.

Thus, \cite{significant} proposed a test method called Significance Test for Lasso. Below, we assume as a null hypothesis that $M_{k-1}$ contains all the indices of the non-zero components of the true $\beta$, denoted $\beta^*$.

In Lasso, let $M_{k-1}$ be the active set just before $\lambda_k$. That is, at $\lambda=\lambda_k$, among $\hat{\beta}(\lambda_k) \in \mathbb{R}^p$, the components of $M_{k-1}$ are non-zero, and
\begin{equation}\label{eq701}
\frac{1}{2}\|y-X\beta\|_2^2+\lambda_k\|\beta\|_1
\end{equation}
is minimized by $\beta=\hat{\beta}(\lambda_k)$, in which the component of the newly selected $j_k \in M_k\backslash M_{k-1}$ is also 0 at $\lambda=\lambda_k$. Assuming $M_{k-1}\subsetneq M_k$, similarly at $\lambda=\lambda_{k+1}$, among $\hat{\beta}(\lambda_{k+1})\in \mathbb{R}^p$, the components of $M_k$ are non-zero, and
\[
\frac{1}{2}\|y-X\beta\|_2^2+\lambda_{k+1}\|\beta\|_1
\]
is minimized by $\beta=\hat{\beta}(\lambda_{k+1})$, in which the component of $j_{k+1} \in M_{k+1}\backslash M_k$ is 0 at $\lambda=\lambda_{k+1}$.

\cite{significant} derived the distribution under the null hypothesis of the statistic
\begin{equation}\label{eq812}
T_k:=\left(\langle y, X\hat{\beta}(\lambda_{k+1})\rangle -\langle y, X_{M_{k-1}}\tilde{\beta}_{M_{k-1}}(\lambda_{k+1})\rangle\right)/\sigma^2
\end{equation}
to test for a significant difference between this $\hat{\beta}(\lambda_{k+1})\in \mathbb{R}^p$ and $\tilde{\beta}=\tilde{\beta}_{M_{k-1}}(\lambda_{k+1})\in \mathbb{R}^{k-1}$, which minimizes
\begin{equation}\label{eq702}
\frac{1}{2}\|y-X_{M_{k-1}}\tilde{\beta}\|_2^2+\lambda_{k+1}\|\tilde{\beta}\|_1.
\end{equation}

In general, in Lasso, there is no guarantee that the variable set $M$ will expand monotonically even if the value of $\lambda$ is decreased. Under $M_{k-1}\subsetneq M_k$, \cite{significant} set
\begin{equation}\label{eq81}
\omega_k=\left\|(X_{M_{k}}^+)^\top s_{M_{k}} - (X_{M_{k-1}}^+)^\top s_{M_{k-1}} \right\|_2
\end{equation}
and derived the following.

\begin{hodai}\label{hodai7}
\begin{equation}\label{eq82}
T_k=\omega_{k}^2\lambda_k(\lambda_k-\lambda_{k+1})/\sigma^2.
\end{equation}
\end{hodai}
(For a proof, please refer to the appendix.)

In the following, we discuss the case where \(X^\top X = I_p\). For a generalization to the general case, please refer to \cite{significant}. In the case of \(X^\top X = I_p\),
\begin{eqnarray*}
\omega_k &=& \left\|(X_{M_{k}}^+)^\top s_{M_{k}} - (X_{M_{k-1}}^+)^\top s_{M_{k-1}} \right\|_2 \\
&=& \left\|X_{M_{k}} s_{M_{k}} - X_{M_{k-1}} s_{M_{k-1}} \right\|_2 = \|s_{j_k}X_{j_k}\| \\
&=& \|X_{j_k}\| = 1,
\end{eqnarray*}
thus,
\begin{equation}\label{eq1241}
T_k = \lambda_k(\lambda_k - \lambda_{k+1})/\sigma^2
\end{equation}
holds. Also, when we let \(U_j = X_j^\top y\), we can write \(\lambda_j = |U_j|\) where \(j = 1, \ldots, p\). However, \(U_1, \ldots, U_p\) are random variables satisfying \(|U_1| \geq |U_2| \geq \ldots \geq |U_p|\). In this case,
\[T_k = |U_k|(|U_k| - |U_{k+1}|)/\sigma^2\]
can be written.

On the other hand, when we consider \(V_1 \geq V_2 \geq \ldots \geq V_p\) as \(p\) independent random variables following a \(\chi_1\) distribution, and when fixing \(r \leq p\), as \(p \rightarrow \infty\),
\begin{equation}\label{eq-exp}
\left(V_1(V_1-V_2), V_2(V_2-V_3), \ldots, V_r(V_r-V_{r+1})\right) \xrightarrow{d} \left(Exp(1),Exp(1/2),\ldots,Exp(1/r)\right)
\end{equation}
is derived (\cite{significant}). Here, \(\xrightarrow{d}\) denotes convergence in distribution, and \(Exp(\alpha)\) is an exponential distribution such that \(Z \sim Exp(\alpha) \Longrightarrow \mathbb{E}[Z] = \alpha\).

Using the statistic \(T_k\), assuming Lasso variable selection and denoting \(M_*\) as the true variable set, a test is constructed that takes the current selected variable set \(M_k\) containing \(M^*\) as a subset as the null hypothesis \(H_0\). If all necessary variables are selected, the unselected variables are merely noise, and it is assumed to follow the null hypothesis.

\cite{significant} specifically proved the following proposition, and constructed a statistical test based on this null distribution.

\begin{prop}[\cite{significant}]
Under the assumption that 
\[
\lim_{{p \to \infty}} \left( \min_{{j \in M^*}} |\beta_j^*| - \sigma\sqrt{2\log p} \right) = \infty,
\]
the following hold as \(p \to \infty\):
\begin{enumerate}
\item The probability of the event \(\min_{{j \in M^*}} |U_j| > \max_{{j \notin M^*}} |U_j|\) converges to 1.
\item \(\left(T_{k_* + 1}, T_{k_* + 2}, \ldots, T_{k_* + r}\right) \xrightarrow{d} \left(Exp(1), Exp\left(\frac{1}{2}\right), \ldots, Exp\left(\frac{1}{r}\right)\right)\).
\end{enumerate}
\end{prop}

\noindent \textbf{Proof of Proposition 1:}\\
Let \(\theta := \min_{{j \in M^*}} \beta_j^*\). By assumption, there exists \(c_p\) such that both \(c_p - \sigma\sqrt{2\log p}\) and \(\theta - c_p\) tend to infinity as \(p \to \infty\). Since \(X^\top X = I_p\), and \(U_j \sim N(\beta_j^*, \sigma^2)\) occur independently, for each \(j \in M^*\), we have
\[
{\mathbb P}(|U_j| \leq c_p) = \Phi\left(\frac{c_p - \beta_j^*}{\sigma}\right) - \Phi\left(\frac{-c_p - \beta_j^*}{\sigma}\right) \leq \Phi\left(\frac{c_p - \beta_j^*}{\sigma}\right) \to 0,
\]
\[
{\mathbb P}\left(\min_{{j \in M^*}}|U_j| > c_p\right) = \prod_{{j \in M^*}}{\mathbb P}(|U_j| > c_p) \to 1,
\]
\[
{\mathbb P}\left(\min_{{j \notin M^*}}|U_j| \leq c_p\right) = \left(1 - 2\Phi^c\left(\frac{c_p}{\sigma}\right)\right)^{p - k_*} \to 1,
\]
where \(\beta_j^* = 0\), \(j \notin M^*\), and
\[
\Phi^c(u) = \frac{1}{\sqrt{2\pi}}\int_u^\infty e^{-x^2/2}dx < \frac{1}{\sqrt{2\pi}}\int_u^\infty \frac{x}{u}e^{-x^2/2}dx = \frac{e^{-u^2/2}}{u\sqrt{2\pi}}, \quad u > 0,
\]
\[
1 - 2\Phi^c(\sqrt{2\log p}) > 1 - \frac{2e^{-\log p}}{\sqrt{2\pi \log p}} = 1 - \frac{2}{p\sqrt{2\pi \log p}},
\]
\[
\left(1 - 2\Phi^c(\sqrt{2\log p})\right)^{p - k_*} > \left\{\left(1 - \frac{2}{p\sqrt{2\pi \log p}}\right)^{\sqrt{2\pi}\, p\sqrt{\log p}}\right\}^{\frac{p-k_*}{p\sqrt{2\pi \log p}}} \to 1.
\]
Let's denote the event \(\min_{{j \in M^*}}|U_j| > \max_{{j \notin M^*}}|U_j|\) by \(B\). Then \({\mathbb P}(B) \to 1\), and therefore, for any event \(E\), \({\mathbb P}(E|B) \to {\mathbb P}(E)\). Under event \(B\), \(T_{k_* + i} = V_i(V_i - V_{i + 1})\), so when \({\mathbb P}(B) \to 1\), \(T_{k_* + i}\) converges in probability to \(V_i(V_i - V_{i + 1})\). From (\ref{eq-exp}), this implies the proposition.\\
\hfill \(\square\)

\section{Spacing Test for LARS}

Lastly, we introduce a simplified Spacing Test, discussed with respect to the selective inference for Lasso in Section 6 (\cite{ryan_exact}), and discuss its relation with the Significance test in Section 7 (\cite{significant}). Here, \cite{ryan_exact} constructed a framework that uniformly handles sequential model selection, including not only Lasso but also Forward Selection (FS) and others. In order to uniformly deal with such sequential model selection, without assuming linear regression and instead letting
\[ y = \theta + \epsilon, \quad \epsilon \sim N(0, \sigma^2I_n), \]
without setting \(\theta = X\beta^*\), \(X \in {\mathbb R}^{n \times p}\), \(\beta^* \in {\mathbb R}^p\), it considered a test regarding the null hypothesis \(v^\top \theta = 0\) for some \(v \in {\mathbb R}^n\). Regarding LARS, \(v\) is set as \(v := (X_{M_{k-1}}^+)^\top e_k\), and \(\theta\) is used as the mean of \(y\) under the null hypothesis, which is \(X\beta^*\). Under the null hypothesis, the \(k-1\) components of \(\beta^*\) are non-zero, while the rest are zero; hence, \(v^\top \theta\) becomes the \(k\)-th component of \(\beta^*\), i.e., \(\beta_{j_k}^*\). That is, the Spacing Test, like the Significance Test, uses a null hypothesis that, given the variables selected so far, the coefficients of any further selected variables become zero (there is no need to select more variables).

Below, we present the polyhedron, null hypothesis, and truncated distribution.

First, we use the following fact.
\begin{hodai}[\cite{ryan_exact}]\label{hodai2}
Equations (\ref{eq43}) and (\ref{eq44}) are equivalent to the following four conditions:
\begin{equation}\label{eq70}
c_1(j_1,s_1)^\top y\geq c_2(j_2,s_2)^\top y \geq \cdots \geq c_k(j_k,s_k)^\top y\geq 0
\end{equation}
\begin{equation}\label{eq72}
c_k(j_k,s_k)^\top y\geq \displaystyle c_{k+1}^*:=\max_{(j,s)\in S^+_k}c_{k+1}(j,s) ^\top y
\end{equation}
\begin{equation}\label{eq73}
c_l(j_l,s_l)^\top y\geq \displaystyle \min_{(j,s)\in S^-_l}c_{l+1}(j,s) ^\top y\ ,\ l=1,\ldots,k
\end{equation}
\begin{equation}\label{eq74}
c_l(j_l,s_l)^\top y\geq \displaystyle \max_{(j,s)\in S^0_l}c_k(j,s) ^\top y\ ,\ l=1,\ldots,k
\end{equation}
where we have set:
\[ S_k^+ := \{(j,s)|j\not\in M_k, c_k(j,s)^\top c_k(j_k,s_k)\leq \|c_k(j_k,s_k)\|_2^2, c_k(j,s)^\top y \leq  c_k(j_k,s_k)^\top y\} \]
\[ S_l^- := \{(j,s)|j\not\in M_l, c_l(j,s)^\top c_l(j_l,s_l)\geq \|c_l(j_l,s_l)\|_2^2, c_l(j,s)^\top y \leq  c_l(j_l,s_l)^\top y\} \]
\[ S_l^0 := \{(j,s)|j\not\in M_l, c_l(j,s)^\top c_l(j_l,s_l)= \|c_l(j_l,s_l)\|_2^2, c_l(j,s)^\top y \leq  c_l(j_l,s_l)^\top y\} \]
\end{hodai}
(See the appendix for proof.)

Furthermore, it has been proposed to approximate equations (\ref{eq70})-(\ref{eq74}) by $k+1$ conditions from equations (\ref{eq70}) and (\ref{eq72}). In fact, when \(X^\top X=I_p\), \(c_l(j,s)=X_j\), and for \(j\not\in M_l\), \(c_l(j,s)^\top c_l(j_l,s_l)=X_j^\top X_l=0\) holds. This means that both \(S_l^-\) and \(S_l^0\) are empty sets and equations (\ref{eq73}) and (\ref{eq74}) are unconditionally satisfied. In the general situation where \(X^\top X\not=I_p\), it becomes an accurate approximation of the polyhedron, but computationally advantageous as the matrix \(A\) has significantly fewer rows. In practice, the constraint conditions in equations (\ref{eq73}) and (\ref{eq74}) are often not applied, and deleting them does not significantly alter the geometric characteristics of the set.

First, we define the row vectors \(A_1,\dots,A_{k+1}\) of matrix \(A\) as
\[ A_l=-c_l(j_l,s_l)^\top+c_{l+1}(j_{l+1},s_{l+1})^\top\ ,\ l=1,\ldots,k-1 \]
\[ A_{k}=A_{k+1}=-c_k(j_k,s_k)^\top \]
and let \(b_1=\ldots = b_k=0\), \(b_{k+1}=c_{k+1}^*\). Constructing the polyhedron \(A_ly\leq b_l\), \(l=1,\ldots,k+1\), and under the null hypothesis \(\beta_{j_k}=0\), the distribution followed by the statistic
\begin{equation}\label{eq301}
T_k:=\frac{\displaystyle \Psi\left(\lambda_{k-1}\frac{\omega_k}{\sigma}\right)-\Psi\left(\lambda_{k}\frac{\omega_k}{\sigma}\right)}
{\displaystyle \Psi\left(\lambda_{k-1}\frac{\omega_k}{\sigma}\right)-\Psi\left({c^*_{k+1}}^\top y\frac{\omega_k}{\sigma}\right)}
\end{equation}
(spacing test) was derived. Here, \(\omega_k\) is given in (\ref{eq81}), and \(\eta^\top y=\lambda_k\) is divided by \(\sigma\) and multiplied by \(\omega_k\) because \(\omega_k\) is the reciprocal of \(\|\eta\|_2\), and to normalize the magnitude of \(\eta\in {\mathbb R}^n\) so that \(\eta^\top y/(\sigma\|\eta\|_2)\) follows the standard normal distribution.

\begin{hodai}\label{hodai3}
\(\|\eta\|_2=\omega_k^{-1}\)
\end{hodai}
(See the appendix for proof.)

\begin{prop}
Assuming the null hypothesis \(H_0: \beta_{j_k}=0\), the conditional probability of event \(\{T_k^*\leq \alpha\}\) under the polyhedron \(\{Ay\leq b\}\) is \(\alpha\).
\begin{equation}\label{eq302}
{\mathbb P}_{H_0}(T_k\leq \alpha|Ay\leq b)=\alpha
\end{equation}
\end{prop}
For a two-sided test, use \(2\min \{T_k,1-T_k\}\) instead of \(T_k\).

\textbf{Proof of Proposition 2:} 
Throughout, to simplify the notation, we write \(c_l:=c_l(j_l,s_l)\), \(l=1,\ldots,k\). First, let \(\eta:=c_k\). If \(h<k\), then
\begin{align}
(P_{h-1}^\perp X_{j_h})^\top P_{k-1}^\perp X_{j_k} &= X_{j_h}^\top P_{h-1}^\perp P_{k-1}^\perp X_{j_k} \nonumber\\
&= X_{j_h}^\top P_{k-1}^\perp  X_{j_k}=0
\end{align}
implying that \(c_h^\top c_k=0\) and, since \(c=\eta/\|\eta\|_2^2\), we get
\[ A_1c=\cdots=A_{k-2}c=0,\]
\[ A_{k-1}c=1, \]
\[ A_kc=A_{k+1}c=-1.\]
Therefore, there is only one \(j\) such that \((Ac)_{j}>0\). Also, there are two \(j\) such that \((Ac)_{j}<0\), but since \(A_{k}=A_{k+1}=-c_k\), \(b_{k}=0\), and \(b_{k+1}=c^{*}_{k+1}\), from equation (\ref{eq91}) we have
\begin{align}
\nu^+(z)&=\frac{b_{k-1}-A_{k-1}z}{A_{k-1}c}\nonumber\\
&=0-\{-c_{k-1}^\top+c_k^\top\}(I_n-\frac{\eta\eta^\top}{\|\eta\|_2^2})y\nonumber\\
&=(c_{k-1}^\top-c_k^\top +c_k^\top \frac{c_kc_k^\top}{\|c_k\|_2^2})y=\lambda_{k-1}
\end{align}
and
\begin{align}
\nu^-(z)&=\frac{b_{k+1}-A_{k+1}z}{A_{k+1}c}\nonumber\\
&=c_{k+1}^*-(-c_k^\top)\left(I_n-\frac{\eta\eta^\top}{\|\eta\|_2^2}\right)y\nonumber\\
&=c_{k+1}^*-(-c_k^\top)\left(I_n-\frac{c_kc_k^\top}{\|c_k\|_2^2}\right)y=c_{k+1}^*
\end{align}
are valid, and the value in equation (\ref{eq19}) is uniformly distributed in \([a,b]=[\nu^-(z),\nu^+(z)]\). The value of \(z_0\) at that time is
\[ z_0=\left(I_n-\frac{c_kc_k^\top}{\|c_k\|_2^2}\right)y. \]
\hfill \(\square\)

\cite{ryan_exact} proposes an approximation due to the complexity of obtaining the value of ${c_{k+1}^*}^\top y$ in equation (\ref{eq301}), which slightly weakens the detection power, as
\[ {T}^{sp}_k:=
\frac{\displaystyle \Psi\left(\lambda_{k-1}\frac{\omega_k}{\sigma}\right)
-\Psi\left(\lambda_{k}\frac{\omega_k}{\sigma}\right)}{\displaystyle \Psi\left(\lambda_{k-1}\frac{\omega_k}{\sigma}\right)
-\Psi\left(\lambda_{k+1}\frac{\omega_k}{\sigma}\right)}
\]
and proves that
\begin{equation}\label{eq303}
{\mathbb P}_{H_0}({T}_k^{sp}\leq \alpha|Ay\leq b)\leq\alpha.
\end{equation}
Because 
\[
\lambda_{k+1}=\max_{(j,s):j\not\in M_{k}}c_{k+1}(j,s)\geq \max_{(j,s)\in S_k^+}c_{k+1}(j,s)=c_{k+1}^*
\]
and \(T_k\leq {T}^{sp}_k\), equation (\ref{eq303}) holds. Furthermore, under the assumption that \(\omega_k\lambda_{k+1}\xrightarrow{P}\infty \) and \(\omega_k^2\lambda_{k-1}(\lambda_{k-1}-\lambda_k)\xrightarrow{P}\infty\), it was demonstrated that there exists a relationship
\begin{equation}
-\log {T}^{sp}_k-T_k^{sig}\xrightarrow{P}0,
\end{equation}
where \(T_k^{sig}\) is the \(T_k\) in equation (\ref{eq812}) for the Significance Test in Lasso, and \(\xrightarrow{P}\) denotes convergence in probability.

\section{Conclusion}
We have discussed selective inference in sparse estimation above. 
There are several explanations about sparse estimation, and one might feel that it is somewhat ``old news.'' In this paper, we tried to follow the conservative mainstream flow (Stanford Statistics) that includes selective inference, Lasso, FS, and LARS. The Significance Test in Section 7 \cite{significant} and the Spacing Test in Section 8 \cite{ryan_exact} are considered to be difficult to understand (the latter presupposes an understanding of the former). And while they are highly cited\footnote{As of March 1, 2023, their citation counts on Google Scholar are 812 and 506, respectively.}, hardly any follow-up research has emerged even after a decade. The honest motive for writing this paper is a belief that if readers can grasp the essence of them through this commentary, it might lead to research opportunities, suspecting that there might be ``buried treasure" that remains unearthed.

Although the number of theoretical papers is decreasing, results applying to specific problems in machine learning and other fields have been presented. Applications to clustering \cite{witten}, to estimating the MMD (maximum mean discrepancy) for assessing differences in distributions \cite{yamada19}, to HSIC-Lasso applications \cite{yamada18}, and others like \cite{takeuchi} mentioned in Section 4, have been presented at top conferences in machine learning. Furthermore, \cite{tasaka} proposes a method for determining the value of \(\lambda\) in Fused Lasso by utilizing the Spacing Test.

\appendix
\numberwithin{equation}{section}
\makeatletter 
\newcommand{\section@cntformat}{Appendix \thesection:\ }

\section{Appendix}

\subsection*{Proof of Lemma 1}
Since \(F_\mu(x)\) is a truncated normal distribution, when the probability density function is denoted as \(f_\mu(x)\), regarding the likelihood ratio,
\[
\mu < \lambda, y < z \Longrightarrow \frac{f_\lambda(z)}{f_\mu(z)}> \frac{f_\lambda(y)}{f_\mu(y)}
\]
holds. Indeed, taking the logarithm on both sides and subtracting the right side from the left side, we get
\[
-\frac{(z-\lambda)^2}{2\sigma^2}+\frac{(z-\mu)^2}{2\sigma^2}+\frac{(y-\lambda)^2}{2\sigma^2}-\frac{(y-\mu)^2}{2\sigma^2}=\frac{(z-y)(\lambda-\mu)}{\sigma^2}>0
\]
At this time, by integrating both sides of \(f_\lambda(z)f_\mu(y)>f_\lambda(y)f_\mu(z)\), we have
\[
f_\lambda(z)F_{\mu}(x)=\int_{-\infty}^x f_\lambda(z)f_\mu(y)dy>\int_{-\infty}^x f_\lambda(y)f_\mu(z)dy=f_\mu(z)F_{\lambda}(x)
\]
\[
F_{\mu}(x)(1-F_\lambda(x))=\int^{\infty}_x f_\lambda(z)F_{\mu}(x)dz>\int^{\infty}_x f_\mu(z)F_{\lambda}(x)dz=F_{\lambda}(x)(1-F_\mu(x))
\]
This implies that \(F_{\mu}(x)>F_{\lambda}(x)\).

\subsection*{Proof of Lemma \ref{hodai-new}}
From (\ref{eq322}) and (\ref{eq321}), for \(j \not\in M_{k-1}\), we can transform
\begin{align*}
X_{j}^\top r_k=s_k\lambda_k 
&\Longleftrightarrow X_{j}^\top \left\{r_{k-1}-\left(1-\frac{\lambda_k}{\lambda_{k-1}}\right)P_{k-1}r_{k-1}\right\}=s_k\lambda_k\\
&\Longleftrightarrow X_{j}^\top P_{k-1}^\perp r_{k-1}=
\lambda_k\left\{s_k-X_j^\top (X_{M_{k-1}}^+)^\top s_{M_{k-1}}\right\} \\
&\Longleftrightarrow \frac{X_{j}^\top P_{k-1}^\perp}{s_k-X_j^\top (X_{M_{k-1}}^+)^\top s_{M_{k-1}}}y=\lambda_k
\end{align*}
In the last transformation, we used
\[
r_k=P_{k-1}^\perp r_{k-1}+\frac{\lambda_k}{\lambda_{k-1}}P_{k-1}r_{k-1}
\]
\[
P_{k-1}^\perp r_{k-1}= P_{k-1}^\perp P_{k-2}^\perp r_{k-2}= P_{k-1}^\perp r_{k-2}=\cdots =P_{k-1}^\perp y
\]

\subsection*{Proof of Lemma \ref{hodai7}}
From (\ref{eq701}), the component of \(\hat{\beta}(\lambda_k)\) corresponding to \(M_{k-1}\) is non-zero. By setting the derivative of
$$\frac{1}{2}\|y-X_{M_{k-1}}\beta\|_2^2+\lambda_k\|\beta\|_1$$
to zero, we have
$$-X_{M_{k-1}}^\top (y-X_{M_{k-1}}\beta)+s_{M_{k-1}}\lambda_k=0$$
for some \(\beta\in \mathbb{R}^{k-1}\). Therefore, using this \(\beta\) we can write
$$X\hat{\beta}(\lambda_k)=X_{M_{k-1}}\beta=X_{M_{k-1}}(X_{M_{k-1}}^\top X_{M_{k-1}})^{-1}(X_{M_{k-1}}^\top y-\lambda_ks_{M_{k-1}})
=P_{k-1}y-\lambda_k(X_{M_{k-1}^\top})^+s_{M_{k-1}}.$$
Similarly,
$$X\hat{\beta}(\lambda_{k+1})=P_{k}y-\lambda_{k+1}(X_{M_{k}^\top})^+s_{M_{k}}$$
holds. Also, the component of \(\tilde{\beta}_{M_{k-1}}(\lambda_{k+1})\) corresponding to \(M_{k-1}\) is non-zero, and for some \(\tilde{\beta}\in \mathbb{R}^{k-1}\) satisfying
$$\frac{1}{2}\|y-X_{M_{k-1}}\tilde{\beta}\|_2^2+\lambda_{k+1}\|\tilde{\beta}\|_1,$$
when differentiated and set to zero, we have
$$X_{M_{k-1}}\tilde{\beta}_{M_{k-1}}(\lambda_{k+1})=P_{k-1}y-\lambda_{k+1}(X_{M_{k-1}^\top})^+s_{M_{k-1}}.$$
Substituting these into the definition of \(T_k\) (\ref{eq812}), we obtain the following equation:
\begin{equation}\label{eq707}
T_k=y^\top (P_k-P_{k-1})y/\sigma^2 -\lambda_{k+1} y^\top\left\{ (X_{M_k}^\top)^+s_{M_k}- (X_{M_{k-1}}^\top)^+s_{M_{k-1}} \right\}/\sigma^2.
\end{equation}
On the other hand, when \(f_k(\lambda):=P_{k}y-\lambda(X_{M_k}^\top)^+s_{M_k}\), by continuity of the Lasso solution path, we have \(f_{k-1}(\lambda_k)=f_k(\lambda)\), i.e.,
$$
P_{k-1}y-\lambda_k(X_{M_{k-1}}^\top)s_{M_{k-1}}=P_{k}y-\lambda_k(X_{M_{k}}^\top)s_{M_{k}}.
$$
Therefore,
\begin{equation}\label{eq708}
(P_k-P_{k-1})y=\lambda_k\left((X_{M_{k}}^+)^\top s_{M_{k}} - (X_{M_{k-1}}^+)^\top s_{M_{k-1}} \right)
\end{equation}
\begin{equation}\label{eq709}
y^\top (P_k-P_{k-1})y=\lambda_k^2\left\|(X_{M_{k}}^+)^\top s_{M_{k}} - (X_{M_{k-1}}^+)^\top s_{M_{k-1}} \right\|_2^2
\end{equation}
hold, where we used \((P_k-P_{k-1})^2=P_k-P_{k-1}\). Finally, taking the inner product of both sides of (\ref{eq708}) with \(y\) and using (\ref{eq709}), we obtain
\begin{equation}\label{eq710}
y^\top \left((X_{M_{k}}^+)^\top s_{M_{k}} - (X_{M_{k-1}}^+)^\top s_{M_{k-1}} \right)
=\lambda_k\left\|(X_{M_{k}}^+)^\top s_{M_{k}} - (X_{M_{k-1}}^+)^\top s_{M_{k-1}} \right\|_2^2
\end{equation}
Substituting (\ref{eq709}) and (\ref{eq710}) into (\ref{eq707}), the lemma is obtained.

\subsection*{Proof of Lemma \ref{hodai2}}
$$h_k(j,s):=
\frac{\displaystyle c_k(j,s)-\frac{c_k(j,s)^\top c_k(j_k,s_k)}{\|c_k(j_k,s_k)\|_2^2}c_{k}(j_k,s_k)}
{\displaystyle 1-\frac{c_k(j,s)^\top c_k(j_k,s_k)}{\|c_k(j_k,s_k)\|_2^2}}$$
We have that,
\begin{eqnarray*}
&&c_k(j_k,s_k)^\top y\geq c_k(j,s)^\top y\\
&\Longleftrightarrow&
c_k(j_k,s_k)^\top 
y\left\{1-\frac{c_k(j,s)^\top c_k(j_k,s_k)}{\|c_k(j_k,s_k)\|_2^2}\right\}
\geq c_k(j,s)^\top y-\frac{c_k(j,s)^\top c_k(j_k,s_k)}{\|c_k(j_k,s_k)\|_2^2}c_k(j_k,s_k)^\top y\\
&\Longleftrightarrow&
\left\{
\begin{array}{ll}
c_k(j_k,s_k)^\top y\geq h_k(j,s)^\top y,&c_k(j,s)^\top c_k(j_k,s_k)\leq\|c_k(j_k,s_k)\|_2^2\\
c_k(j_k,s_k)^\top y\leq h_k(j,s)^\top y,&c_k(j,s)^\top c_k(j_k,s_k)\geq \|c_k(j_k,s_k)\|_2^2\\
c_k(j_k,s_k)^\top y\geq  c_k(j,s)^\top y,& c_k(j,s)^\top c_k(j_k,s_k)=\|c_k(j_k,s_k)\|_2^2\\
\end{array}
\right.
\end{eqnarray*}
Therefore,
\begin{eqnarray}
&&c_k(j_k,s_k)^\top y\geq c_k(j,s)^\top y\ ,\ (j,s)\not=(j_k,s_k)\nonumber\\
&\Longleftrightarrow&
\left\{
\begin{array}{l}
c_k(j_k,s_k)^\top y\geq  c_k(j_k,-s_k)^\top y\Longleftrightarrow c_k(j_k,s_k)^\top y\geq  0\\
c_k(j_k,s_k)^\top y\geq \displaystyle \max_{c_k(j,s)^\top c_k(j_k,s_k)\leq\|c_k(j_k.s_k)\|_2^2}h_k(j,s) ^\top y\\
c_k(j_k,s_k)^\top y\leq\displaystyle \min_{c_k(j,s)^\top c_k(j_k,s_k)\geq \|c_k(j_k.s_k)\|_2^2}h_k(j,s) ^\top y\\
c_k(j_k,s_k)^\top y\geq \displaystyle \max_{c_k(j,s)^\top c_k(j_k,s_k)=\|c_k(j_k.s_k)\|_2^2}c_k(j,s) ^\top y\\
\end{array}
\right.
\end{eqnarray}
holds (\cite{significant}). Next, we consider the following lemma.
\begin{hodai}[\cite{significant}]\label{hodai4}
$$
h_k(j,s)=c_{k+1}(j,s)$$
\end{hodai}
(The proof follows after the proof of Lemma \ref{hodai3})

Consequently, equations (\ref{eq41}), (\ref{eq43}), and (\ref{eq44}) are equivalent to the following five conditions \cite{ryan_exact}:
\begin{equation}\label{eq80-1}
c_l(j_l,s_l)^\top y\leq \lambda_{l-1} ,\ l=1,\ldots,k
\end{equation}
\begin{equation}\label{eq81-1}
c_l(j_l,s_l)^\top y \geq 0 ,\ l=1,\ldots,k
\end{equation}
\begin{equation}\label{eq82-1}
c_l(j_l,s_l)^\top y\geq \displaystyle \max_{(j,s)\in S^+_l}c_{l+1}(j,s) ^\top y ,\ l=1,\ldots,k
\end{equation}
\begin{equation}\label{eq83-1}
c_l(j_l,s_l)^\top y\geq \displaystyle \min_{(j,s)\in S^-_l}c_{l+1}(j,s) ^\top y\ ,\ l=1,\ldots,k
\end{equation}
\begin{equation}\label{eq84-1}
c_l(j_l,s_l)^\top y\geq \displaystyle \max_{(j,s)\in S^0_l}c_k(j,s) ^\top y\ ,\ l=1,\ldots,k
\end{equation}
First, since $\lambda_l=c_l(j_l,s_l)^\top y\geq c_{l+1}(j_{l+1},s_{l+1})^\top y$, equations (\ref{eq80-1}) and (\ref{eq81-1}) are equivalent to equation (\ref{eq70}). Additionally, in general
$$c_{l}(j_l,s_l)\geq c_{l}(j_{l+1},s_{l+1})=\max_{(j,s):j\not\in M_l}c_{l+1}(j,s) ^\top y\geq \max_{(j,s)\in S^+_l}c_{l+1}(j,s) ^\top y$$
therefore, condition (\ref{eq82-1}) can be omitted for $l=1,\ldots,k-1$. Thus, the proposition is obtained.

\subsection*{Proof of Lemma \ref{hodai3}}
\begin{eqnarray*}
&&X_{M_{k}}^\top X_{M_{k}}
\left[
\begin{array}{c}
z_{1}\\
z_{2}
\end{array}
\right]
=
\left[
\begin{array}{cc}
X_{M_{k-1}}^\top X_{M_{k-1}}& X_{M_{k-1}}^\top X_{j_k}\\
X_{j_k}^\top X_{M_{k-1}}& X_{j_k}^\top X_{j_k}
\end{array}
\right]
\left[
\begin{array}{c}
z_{1}\\
z_{2}
\end{array}
\right]
=
\left[
\begin{array}{c}
s_{M_{k-1}}\\
s_{j_k}
\end{array}
\right]\\
&\Longleftrightarrow&
\left\{
\begin{array}{l}
z_1=(X_{M_{k-1}}^\top X_{M_{k-1}})^{-1}s_{M_{k-1}}- X_{M_{k-1}}^+ X_{j_k}z_2\\
\displaystyle z_2=\frac{s_{k}-s_{M_{k-1}}X_{M_{k-1}}^+ X_{j_k}}{X_{j_k}^\top P^\perp_{k-1}X_{j_{k}}}
\end{array}
\right.
\end{eqnarray*}
Hence, the following transformation can be made:
\begin{eqnarray*}
\omega_k^2&=&\|(X_{M_k}^\top)^+s_{M_k}-(X_{M_{k-1}}^\top)^+s_{M_{k-1}}\|^2_2\\
&=&s_{M_k}^\top (X^\top_{M_k} X_{M_k})s_{M_k} -s_{M_{k-1}}^\top (X^\top_{M_{k-1}} X_{M_{k-1}})s_{M_{k-1}}\\
&=&s_{M_{k-1}}^\top z_1+ s_{k}^\top z_2 -s_{M_{k-1}}^\top (X^\top_{M_{k-1}} X_{M_{k-1}})s_{M_{k-1}}\\
&=&[s_k-s_{M_{k-1}}^\top X_{M_{k-1}}^+X_{j_k}]z_2\\
&=&\frac{\{s_{k}-s_{M_{k-1}}X_{M_{k-1}}^+ X_{j_k}\}^2}{X_{j_k}^\top P^\perp_{k-1}X_{j_{k}}}
= \|\eta\|_2^{-2}
\end{eqnarray*}

\subsection*{Proof of Lemma \ref{hodai4}} 
From (\ref{eq107}), we define
$$\theta_{k,j}:=\frac{X_{j_k}^\top P_{k-1}^\perp X_{j}}{X_{j_k}^\top P_{k-1}^\perp X_{j_k}}$$
and need to show
\begin{equation}\label{eq101}
\frac{X_j^\top P_{k-1}^\perp y-\theta_{k,j}X_{j_k}^\top P_{k-1y}^\perp}{
\{s-s_{M_{k-1}}(X_{M_{k-1}})^+X_j\}-\theta_{k,j}\{s_k-s_{M_{k-1}}(X_{M_{k-1}})^+X_{j_k}\}
}
=
\frac{X_j^\top P_{k}^\perp y}{s-s_{M_{k}}(X_{M_{k}})^+X_j}
\end{equation}
(\cite{significant}). Since $\theta_{k,j}$ is the coefficient of $X_{j_k}$ when $X_{j}$ is the dependent variable and $X_{M_k}$ are the independent variables, $\theta_{k,j}$ becomes the $j_k$-th component of $X_{M_k}^+X_j$. Thus,
\begin{eqnarray*}
&&X_{M_k}^+X_j=[\theta_{M_{k-1},j},\theta_{k,j}]^\top \Longleftrightarrow X_{M_{k-1}}\theta_{M_{k-1},j}+X_{j_k}\theta_{k,j}=P_kX_j\\
&\Longleftrightarrow&
\theta_{M_{k-1},j}=X_{M_{k-1}}^+(P_kX_j-\theta_{k,j}X_{j_k})=X_{M_{k-1}}^+(X_j-\theta_{k,j}X_{j_k})
\end{eqnarray*}
Therefore, the denominator of the left-hand side of (\ref{eq101}) is
$$
s-[s_{M_{k-1}},s_k]^\top \left[
\begin{array}{c}
X_{M_{k-1}}^+(X_j-\theta_{k,j}X_{j_k})\\
\theta_{k,j}
\end{array}
\right]=
s-s_{M_k}^\top \left[
\begin{array}{c}
\theta_{M_{k-1},j}\\
\theta_{k,j}
\end{array}
\right]=s-s_{M_{k}}^\top(X_{M_{k}})^+X_j
$$
which matches the denominator on the right-hand side. Also, since $P_{k-1}^\perp(X_j-\theta_{k,j}X_{j_k})=P_k^\perp X_j$, the numerators on both sides of (\ref{eq101}) match as well.

{\bf Acknowledgments}: We express our gratitude to the handling editor, Professor Hidetoshi Matsui, and two other reviewers, for their numerous comments in great detail. We take this opportunity to extend our thanks.

\frenchspacing
\bibliography{ref_eng}

\end{document}